# Indirect measurement of $\sin^2\theta_W$ (or $M_W$) using $\mu^+\mu^-$ pairs from $\gamma^*/Z$ bosons produced in $p\bar{p}$ collisions at a center-of-momentum energy of 1.96 TeV

T. Aaltonen,[21] S. Amerio[jj,39] D. Amidei,[31] A. Anastassov[v,15] A. Annovi,[17] J. Antos,[12] G. Apollinari,[15] J.A. Appel,[15] T. Arisawa,[52] A. Artikov,[13] J. Asaadi,[47] W. Ashmanskas,[15] B. Auerbach,[2] A. Aurisano,[47] F. Azfar,[38] W. Badgett,[15] T. Bae,[25] A. Barbaro-Galtieri,[26] V.E. Barnes,[43] B.A. Barnett,[23] P. Barria[ll,41] P. Bartos,[12] M. Bauce[jj,39] F. Bedeschi,[41] S. Behari,[15] G. Bellettini[kk,41] J. Bellinger,[54] D. Benjamin,[14] A. Beretvas,[15] A. Bhatti,[45] K.R. Bland,[5] B. Blumenfeld,[23] A. Bocci,[14] A. Bodek,[44] D. Bortoletto,[43] J. Boudreau,[42] A. Boveia,[11] L. Brigliadori[ii,6] C. Bromberg,[32] E. Brucken,[21] J. Budagov,[13] H.S. Budd,[44] K. Burkett,[15] G. Busetto[jj,39] P. Bussey,[19] P. Butti[kk,41] A. Buzatu,[19] A. Calamba,[10] S. Camarda,[4] M. Campanelli,[28] F. Canelli[cc,11] B. Carls,[22] D. Carlsmith,[54] R. Carosi,[41] S. Carrillo[l,16] B. Casal[j,9] M. Casarsa,[48] A. Castro[ii,6] P. Catastini,[20] D. Cauz[qqrr,48] V. Cavaliere,[22] M. Cavalli-Sforza,[4] A. Cerri[e,26] L. Cerrito[q,28] Y.C. Chen,[1] M. Chertok,[7] G. Chiarelli,[41] G. Chlachidze,[15] K. Cho,[25] D. Chokheli,[13] A. Clark,[18] C. Clarke,[53] M.E. Convery,[15] J. Conway,[7] M. Corbo[y,15] M. Cordelli,[17] C.A. Cox,[7] D.J. Cox,[7] M. Cremonesi,[41] D. Cruz,[47] J. Cuevas[x,9] R. Culbertson,[15] N. d'Ascenzo[u,15] M. Datta[ff,15] P. de Barbaro,[44] L. Demortier,[45] M. Deninno,[6] M. D'Errico[jj,39] F. Devoto,[21] A. Di Canto[kk,41] B. Di Ruzza[p,15] J.R. Dittmann,[5] S. Donati[kk,41] M. D'Onofrio,[27] M. Dorigo[ss,48] A. Driutti[qqrr,48] K. Ebina,[52] R. Edgar,[31] A. Elagin,[47] R. Erbacher,[7] S. Errede,[22] B. Esham,[22] S. Farrington,[38] J.P. Fernández Ramos,[29] R. Field,[16] G. Flanagan[s,15] R. Forrest,[7] M. Franklin,[20] J.C. Freeman,[15] H. Frisch,[11] Y. Funakoshi,[52] C. Galloni[kk,41] A.F. Garfinkel,[43] P. Garosi[ll,41] H. Gerberich,[22] E. Gerchtein,[15] S. Giagu,[46] V. Giakoumopoulou,[3] K. Gibson,[42] C.M. Ginsburg,[15] N. Giokaris,[3] P. Giromini,[17] G. Giurgiu,[23] V. Glagolev,[13] D. Glenzinski,[15] M. Gold,[34] D. Goldin,[47] A. Golossanov,[15] G. Gomez,[9] G. Gomez-Ceballos,[30] M. Goncharov,[30] O. González López,[29] I. Gorelov,[34] A.T. Goshaw,[14] K. Goulianos,[45] E. Gramellini,[6] S. Grinstein,[4] C. Grosso-Pilcher,[11] R.C. Group,[51,15] J. Guimaraes da Costa,[20] S.R. Hahn,[15] J.Y. Han,[44] F. Happacher,[17] K. Hara,[49] M. Hare,[50] R.F. Harr,[53] T. Harrington-Taber[m,15] K. Hatakeyama,[5] C. Hays,[38] J. Heinrich,[40] M. Herndon,[54] A. Hocker,[15] Z. Hong,[47] W. Hopkins[f,15] S. Hou,[1] R.E. Hughes,[35] U. Husemann,[55] M. Hussein[aa,32] J. Huston,[32] G. Introzzi[nnoo,41] M. Iori[pp,46] A. Ivanov[o,7] E. James,[15] D. Jang,[10] B. Jayatilaka,[15] E.J. Jeon,[25] S. Jindariani,[15] M. Jones,[43] K.K. Joo,[25] S.Y. Jun,[10] T.R. Junk,[15] M. Kambeitz,[24] T. Kamon,[25,47] P.E. Karchin,[53] A. Kasmi,[5] Y. Kato[n,37] W. Ketchum[gg,11] J. Keung,[40] B. Kilminster[cc,15] D.H. Kim,[25] H.S. Kim,[25] J.E. Kim,[25] M.J. Kim,[17] S.H. Kim,[49] S.B. Kim,[25] Y.J. Kim,[25] Y.K. Kim,[11] N. Kimura,[52] M. Kirby,[15] K. Knoepfel,[15] K. Kondo,[52,*] D.J. Kong,[25] J. Konigsberg,[16] A.V. Kotwal,[14] M. Kreps,[24] J. Kroll,[40] M. Kruse,[14] T. Kuhr,[24] M. Kurata,[49] A.T. Laasanen,[43] S. Lammel,[15] M. Lancaster,[28] K. Lannon[w,35] G. Latino[ll,41] H.S. Lee,[25] J.S. Lee,[25] S. Leo,[41] S. Leone,[41] J.D. Lewis,[15] A. Limosani[r,14] E. Lipeles,[40] A. Lister[a,18] H. Liu,[51] Q. Liu,[43] T. Liu,[15] S. Lockwitz,[55] A. Loginov,[55] D. Lucchesi[jj,39] A. Lucà,[17] J. Lueck,[24] P. Lujan,[26] P. Lukens,[15] G. Lungu,[45] J. Lys,[26] R. Lysak[d,12] R. Madrak,[15] P. Maestro[ll,41] S. Malik,[45] G. Manca[b,27] A. Manousakis-Katsikakis,[3] L. Marchese[hh,6] F. Margaroli,[46] P. Marino[mm,41] M. Martínez,[4] K. Matera,[22] M.E. Mattson,[53] A. Mazzacane,[15] P. Mazzanti,[6] R. McNulty[i,27] A. Mehta,[27] P. Mehtala,[21] C. Mesropian,[45] T. Miao,[15] D. Mietlicki,[31] A. Mitra,[1] H. Miyake,[49] S. Moed,[15] N. Moggi,[6] C.S. Moon[y,15] R. Moore[ddee,15] M.J. Morello[mm,41] A. Mukherjee,[15] Th. Muller,[24] P. Murat,[15] M. Mussini[ii,6] J. Nachtman[m,15] Y. Nagai,[49] J. Naganoma,[52] I. Nakano,[36] A. Napier,[50] J. Nett,[47] C. Neu,[51] T. Nigmanov,[42] L. Nodulman,[2] S.Y. Noh,[25] O. Norniella,[22] L. Oakes,[38] S.H. Oh,[14] Y.D. Oh,[25] I. Oksuzian,[51] T. Okusawa,[37] R. Orava,[21] L. Ortolan,[4] C. Pagliarone,[48] E. Palencia[e,9] P. Palni,[34] V. Papadimitriou,[15] W. Parker,[54] G. Pauletta[qqrr,48] M. Paulini,[10] C. Paus,[30] T.J. Phillips,[14] G. Piacentino,[41] E. Pianori,[40] J. Pilot,[7] K. Pitts,[22] C. Plager,[8] L. Pondrom,[54] S. Poprocki[f,15] K. Potamianos,[26] A. Pranko,[26] F. Prokoshin[z,13] F. Ptohos[g,17] G. Punzi[kk,41] N. Ranjan,[43] I. Redondo Fernández,[29] P. Renton,[38] M. Rescigno,[46] F. Rimondi,[6,*] L. Ristori,[41,15] A. Robson,[19] T. Rodriguez,[40] S. Rolli[h,50] M. Ronzani[kk,41] R. Roser,[15] J.L. Rosner,[11] F. Ruffini[ll,41] A. Ruiz,[9] J. Russ,[10] V. Rusu,[15] W.K. Sakumoto,[44] Y. Sakurai,[52] L. Santi[qqrr,48] K. Sato,[49] V. Saveliev[u,15] A. Savoy-Navarro[y,15] P. Schlabach,[15] E.E. Schmidt,[15] T. Schwarz,[31] L. Scodellaro,[9] F. Scuri,[41] S. Seidel,[34] Y. Seiya,[37] A. Semenov,[13] F. Sforza[kk,41] S.Z. Shalhout,[7] T. Shears,[27] P.F. Shepard,[42] M. Shimojima[t,49] M. Shochet,[11] I. Shreyber-Tecker,[33] A. Simonenko,[13] K. Sliwa,[50] J.R. Smith,[7] F.D. Snider,[15] H. Song,[42] V. Sorin,[4] R. St. Denis,[19] M. Stancari,[15] D. Stentz[v,15] J. Strologas,[34] Y. Sudo,[49] A. Sukhanov,[15] I. Suslov,[13] K. Takemasa,[49] Y. Takeuchi,[49] J. Tang,[11] M. Tecchio,[31] P.K. Teng,[1] J. Thom[f,15] E. Thomson,[40] V. Thukral,[47] D. Toback,[47] S. Tokar,[12] K. Tollefson,[32] T. Tomura,[49] D. Tonelli[e,15] S. Torre,[17] D. Torretta,[15] P. Totaro,[39]

M. Trovato[mm],[41] F. Ukegawa,[49] S. Uozumi,[25] F. Vázquez[l],[16] G. Velev,[15] C. Vellidis,[15] C. Vernieri[mm],[41] M. Vidal,[43] R. Vilar,[9] J. Vizán[bb],[9] M. Vogel,[34] G. Volpi,[17] P. Wagner,[40] R. Wallny[j],[15] S.M. Wang,[1] D. Waters,[28] W.C. Wester III,[15] D. Whiteson[c],[40] A.B. Wicklund,[2] S. Wilbur,[7] H.H. Williams,[40] J.S. Wilson,[31] P. Wilson,[15] B.L. Winer,[35] P. Wittich[f],[15] S. Wolbers,[15] H. Wolfe,[35] T. Wright,[31] X. Wu,[18] Z. Wu,[5] K. Yamamoto,[37] D. Yamato,[37] T. Yang,[15] U.K. Yang,[25] Y.C. Yang,[25] W.-M. Yao,[26] G.P. Yeh,[15] K. Yi[m],[15] J. Yoh,[15] K. Yorita,[52] T. Yoshida[k],[37] G.B. Yu,[14] I. Yu,[25] A.M. Zanetti,[48] Y. Zeng,[14] C. Zhou,[14] and S. Zucchelli[ii6]

(CDF Collaboration)[†]

[1]*Institute of Physics, Academia Sinica, Taipei, Taiwan 11529, Republic of China*

[2]*Argonne National Laboratory, Argonne, Illinois 60439, USA*

[3]*University of Athens, 157 71 Athens, Greece*

[4]*Institut de Fisica d'Altes Energies, ICREA, Universitat Autonoma de Barcelona, E-08193, Bellaterra (Barcelona), Spain*

[5]*Baylor University, Waco, Texas 76798, USA*

[6]*Istituto Nazionale di Fisica Nucleare Bologna, [ii]University of Bologna, I-40127 Bologna, Italy*

[7]*University of California, Davis, Davis, California 95616, USA*

[8]*University of California, Los Angeles, Los Angeles, California 90024, USA*

[9]*Instituto de Fisica de Cantabria, CSIC-University of Cantabria, 39005 Santander, Spain*

[10]*Carnegie Mellon University, Pittsburgh, Pennsylvania 15213, USA*

[11]*Enrico Fermi Institute, University of Chicago, Chicago, Illinois 60637, USA*

[12]*Comenius University, 842 48 Bratislava, Slovakia; Institute of Experimental Physics, 040 01 Kosice, Slovakia*

[13]*Joint Institute for Nuclear Research, RU-141980 Dubna, Russia*

[14]*Duke University, Durham, North Carolina 27708, USA*

[15]*Fermi National Accelerator Laboratory, Batavia, Illinois 60510, USA*

[16]*University of Florida, Gainesville, Florida 32611, USA*

[17]*Laboratori Nazionali di Frascati, Istituto Nazionale di Fisica Nucleare, I-00044 Frascati, Italy*

[18]*University of Geneva, CH-1211 Geneva 4, Switzerland*

[19]*Glasgow University, Glasgow G12 8QQ, United Kingdom*

[20]*Harvard University, Cambridge, Massachusetts 02138, USA*

[21]*Division of High Energy Physics, Department of Physics, University of Helsinki, FIN-00014, Helsinki, Finland; Helsinki Institute of Physics, FIN-00014, Helsinki, Finland*

[22]*University of Illinois, Urbana, Illinois 61801, USA*

[23]*The Johns Hopkins University, Baltimore, Maryland 21218, USA*

[24]*Institut für Experimentelle Kernphysik, Karlsruhe Institute of Technology, D-76131 Karlsruhe, Germany*

[25]*Center for High Energy Physics: Kyungpook National University, Daegu 702-701, Korea; Seoul National University, Seoul 151-742, Korea; Sungkyunkwan University, Suwon 440-746, Korea; Korea Institute of Science and Technology Information, Daejeon 305-806, Korea; Chonnam National University, Gwangju 500-757, Korea; Chonbuk National University, Jeonju 561-756, Korea; Ewha Womans University, Seoul, 120-750, Korea*

[26]*Ernest Orlando Lawrence Berkeley National Laboratory, Berkeley, California 94720, USA*

[27]*University of Liverpool, Liverpool L69 7ZE, United Kingdom*

[28]*University College London, London WC1E 6BT, United Kingdom*

[29]*Centro de Investigaciones Energeticas Medioambientales y Tecnologicas, E-28040 Madrid, Spain*

[30]*Massachusetts Institute of Technology, Cambridge, Massachusetts 02139, USA*

[31]*University of Michigan, Ann Arbor, Michigan 48109, USA*

[32]*Michigan State University, East Lansing, Michigan 48824, USA*

[33]*Institution for Theoretical and Experimental Physics, ITEP, Moscow 117259, Russia*

[34]*University of New Mexico, Albuquerque, New Mexico 87131, USA*

[35]*The Ohio State University, Columbus, Ohio 43210, USA*

[36]*Okayama University, Okayama 700-8530, Japan*

[37]*Osaka City University, Osaka 558-8585, Japan*

[38]*University of Oxford, Oxford OX1 3RH, United Kingdom*

[39]*Istituto Nazionale di Fisica Nucleare, Sezione di Padova, [jj]University of Padova, I-35131 Padova, Italy*

[40]*University of Pennsylvania, Philadelphia, Pennsylvania 19104, USA*

[41]*Istituto Nazionale di Fisica Nucleare Pisa, [kk]University of Pisa, [ll]University of Siena, [mm]Scuola Normale Superiore, I-56127 Pisa, Italy, [nn]INFN Pavia, I-27100 Pavia, Italy, [oo]University of Pavia, I-27100 Pavia, Italy*

[42]*University of Pittsburgh, Pittsburgh, Pennsylvania 15260, USA*

[43]*Purdue University, West Lafayette, Indiana 47907, USA*

[44]*University of Rochester, Rochester, New York 14627, USA*

[45]*The Rockefeller University, New York, New York 10065, USA*

[46] *Istituto Nazionale di Fisica Nucleare, Sezione di Roma 1,*
[pp] *Sapienza Università di Roma, I-00185 Roma, Italy*
[47] *Mitchell Institute for Fundamental Physics and Astronomy,*
*Texas A&M University, College Station, Texas 77843, USA*
[48] *Istituto Nazionale di Fisica Nucleare Trieste,* [qq] *Gruppo Collegato di Udine,*
[rr] *University of Udine, I-33100 Udine, Italy,* [ss] *University of Trieste, I-34127 Trieste, Italy*
[49] *University of Tsukuba, Tsukuba, Ibaraki 305, Japan*
[50] *Tufts University, Medford, Massachusetts 02155, USA*
[51] *University of Virginia, Charlottesville, Virginia 22906, USA*
[52] *Waseda University, Tokyo 169, Japan*
[53] *Wayne State University, Detroit, Michigan 48201, USA*
[54] *University of Wisconsin, Madison, Wisconsin 53706, USA*
[55] *Yale University, New Haven, Connecticut 06520, USA*

Drell-Yan lepton pairs are produced in the process $p\bar{p} \to \mu^+\mu^- + X$ through an intermediate $\gamma^*/Z$ boson. The forward-backward asymmetry in the polar-angle distribution of the $\mu^-$ as a function of the invariant mass of the $\mu^+\mu^-$ pair is used to obtain the effective leptonic determination $\sin^2\theta^{\rm lept}_{\rm eff}$ of the electroweak-mixing parameter $\sin^2\theta_W$, from which the value of $\sin^2\theta_W$ is derived assuming the standard model. The measurement sample, recorded by the Collider Detector at Fermilab (CDF), corresponds to 9.2 fb$^{-1}$ of integrated luminosity from $p\bar{p}$ collisions at a center-of-momentum energy of 1.96 TeV, and is the full CDF Run II data set. The value of $\sin^2\theta^{\rm lept}_{\rm eff}$ is found to be $0.2315 \pm 0.0010$, where statistical and systematic uncertainties are combined in quadrature. When interpreted within the context of the standard model using the on-shell renormalization scheme, where $\sin^2\theta_W = 1 - M_W^2/M_Z^2$, the measurement yields $\sin^2\theta_W = 0.2233 \pm 0.0009$, or equivalently a $W$-boson mass of $80.365 \pm 0.047$ GeV$/c^2$. The value of the $W$-boson mass is in agreement with previous determinations in electron-positron collisions and at the Tevatron collider.

PACS numbers: 12.15.Lk, 13.85.Qk, 14.70.Hp

## I. INTRODUCTION

In this paper, the angular distribution of charged leptons ($\ell^\pm$) from the Drell-Yan [1] process is used to measure the electroweak-mixing parameter $\sin^2\theta_W$ [2]. At the Fermilab Tevatron, Drell-Yan pairs are produced by the process $p\bar{p} \to \ell^+\ell^- + X$, where the $\ell^+\ell^-$ pair is produced through an intermediate $\gamma^*/Z$ boson, and $X$ is the hadronic final state associated with the production of the boson. In the standard model, the production of Drell-Yan lepton pairs at the Born level proceeds through two parton-level processes,

$$q\bar{q} \to \gamma^* \to \ell^+\ell^- \text{ and}$$
$$q\bar{q} \to Z \to \ell^+\ell^-.$$

where the $q$ and $\bar{q}$ are the quark and antiquark, respectively, from the colliding hadrons. The virtual photon couples the vector currents of the incoming and outgoing fermions ($f$), and the spacetime structure of a photon-fermion interaction vertex is $\langle \bar{f}|Q_f\gamma_\mu|f\rangle$, where $Q_f$, the strength of the coupling, is the fermion charge (in units of $e$), and $|f\rangle$ is the spinor for fermion $f$. An interaction vertex of a fermion with a $Z$ boson contains both vector ($V$) and axial-vector ($A$) current components, and its

* Deceased
† With visitors from [a] University of British Columbia, Vancouver, BC V6T 1Z1, Canada, [b] Istituto Nazionale di Fisica Nucleare, Sezione di Cagliari, 09042 Monserrato (Cagliari), Italy, [c] University of California Irvine, Irvine, CA 92697, USA, [d] Institute of Physics, Academy of Sciences of the Czech Republic, 182 21, Czech Republic, [e] CERN, CH-1211 Geneva, Switzerland, [f] Cornell University, Ithaca, NY 14853, USA, [g] University of Cyprus, Nicosia CY-1678, Cyprus, [h] Office of Science, U.S. Department of Energy, Washington, DC 20585, USA, [i] University College Dublin, Dublin 4, Ireland, [j] ETH, 8092 Zürich, Switzerland, [k] University of Fukui, Fukui City, Fukui Prefecture, Japan 910-0017, [l] Universidad Iberoamericana, Lomas de Santa Fe, México, C.P. 01219, Distrito Federal, [m] University of Iowa, Iowa City, IA 52242, USA, [n] Kinki University, Higashi-Osaka City, Japan 577-8502, [o] Kansas State University, Manhattan, KS 66506, USA, [p] Brookhaven National Laboratory, Upton, NY 11973, USA, [q] Queen Mary, University of London, London, E1 4NS, United Kingdom, [r] University of Melbourne, Victoria 3010, Australia, [s] Muons, Inc., Batavia, IL 60510, USA, [t] Nagasaki Institute of Applied Science, Nagasaki 851-0193, Japan, [u] National Research Nuclear University, Moscow 115409, Russia, [v] Northwestern University, Evanston, IL 60208, USA, [w] University of Notre Dame, Notre Dame, IN 46556, USA, [x] Universidad de Oviedo, E-33007 Oviedo, Spain, [y] CNRS-IN2P3, Paris, F-75205 France, [z] Universidad Tecnica Federico Santa Maria, 110v Valparaiso, Chile, [aa] The University of Jordan, Amman 11942, Jordan, [bb] Universite catholique de Louvain, 1348 Louvain-La-Neuve, Belgium, [cc] University of Zürich, 8006 Zürich, Switzerland, [dd] Massachusetts General Hospital, Boston, MA 02114 USA, [ee] Harvard Medical School, Boston, MA 02114 USA, [ff] Hampton University, Hampton, VA 23668, USA, [gg] Los Alamos National Laboratory, Los Alamos, NM 87544, USA,
[hh] Università degli Studi di Napoli Federico I, I-80138 Napoli, Italy

structure is $\langle \bar{f}|g_V^f \gamma_\mu + g_A^f \gamma_\mu \gamma_5 |f\rangle$. The Born-level coupling strengths are

$$g_V^f = T_3^f - 2Q_f \sin^2\theta_W \text{ and}$$
$$g_A^f = T_3^f ,$$

where $T_3^f$ is the third component of the fermion weak isospin, which is $T_3^f = \frac{1}{2}$ $(-\frac{1}{2})$ for positively (negatively) charged fermions. At the Born level, and in all orders of the on-shell renormalization scheme, the $\sin^2\theta_W$ parameter is related to the $W$-boson mass $M_W$ and the $Z$-boson mass $M_Z$ by the relationship $\sin^2\theta_W = 1 - M_W^2/M_Z^2$. Weak-interaction radiative corrections alter the strength of the Born-level couplings into effective couplings. These effective couplings have been investigated at the Tevatron [3–5], at the LHC [6], and at LEP-1 and SLD [7]. Similar couplings have been investigated with neutrino-nucleon collisions at the Tevatron [8] and with electron-proton collisions at HERA [9].

The effective $\sin^2\theta_W$ coupling at the lepton vertex, denoted as $\sin^2\theta_{\text{eff}}^{\text{lept}}$, has been accurately measured at the LEP-1 and SLD $e^+e^-$ colliders. The combined average of six individual measurements yields a value of $0.23153 \pm 0.00016$ [7]. However, there is tension between the two most precise individual measurements: the combined LEP-1 and SLD $b$-quark forward-backward asymmetry ($A_{\text{FB}}^{0,\text{b}}$) yields $\sin^2\theta_{\text{eff}}^{\text{lept}} = 0.23221 \pm 0.00029$, and the SLD polarized left-right asymmetry ($\mathcal{A}_\ell$) yields $\sin^2\theta_{\text{eff}}^{\text{lept}} = 0.23098 \pm 0.00026$. They differ by 3.2 standard deviations.

The Drell-Yan process at hadron-hadron colliders is also sensitive to the $\sin^2\theta_{\text{eff}}^{\text{lept}}$ coupling. Measurements of the forward-backward asymmetry in the $\ell^-$ polar angle distribution as a function of the lepton-pair invariant mass are used to extract the coupling. This paper presents a new measurement of the $\sin^2\theta_{\text{eff}}^{\text{lept}}$ coupling and an inference of the $\sin^2\theta_W$ parameter using a sample of $\mu^+\mu^-$ pairs corresponding to an integrated luminosity of 9.2 fb$^{-1}$ collected at the Tevatron $p\bar{p}$ collider. Innovative methods for the calibration of the muon momentum and measurement of the forward-backward asymmetry are used. Electroweak radiative corrections used for the extraction of $\sin^2\theta_{\text{eff}}^{\text{lept}}$ and $\sin^2\theta_W$ are derived from an approach used at LEP-1 and SLD.

Section II provides an overview of the lepton angular distributions and the extraction of $\sin^2\theta_{\text{eff}}^{\text{lept}}$. Section III discusses quantum chromodynamics (QCD) calculations for the forward-backward asymmetry and the inclusion of electroweak radiative-correction form factors used in the analysis of high energy $e^+e^-$ collisions. These form factors are important in determining $\sin^2\theta_W$ from the measurement of $\sin^2\theta_{\text{eff}}^{\text{lept}}$. Section IV describes the experimental apparatus. Section V reports on the selection of data for the measurement of the forward-backward asymmetry. Section VI describes the simulation of the reconstructed data. Section VII presents the measurement of the asymmetry and the corrections made to the data and simulation. Section VIII describes the method used to extract $\sin^2\theta_{\text{eff}}^{\text{lept}}$. Section IX describes the systematic uncertainties. Finally, Sec. X gives the results, and Sec. XI presents the summary. The units $\hbar = c = 1$ are used for equations and symbols, but standard units are used for numerical values of particle masses and momenta, e.g., 40 GeV/$c^2$ and 20 GeV/$c$, respectively, where $c$ denotes the speed of light.

## II. LEPTON ANGULAR DISTRIBUTIONS

The angular distribution of leptons from the Drell-Yan process in the rest frame of the boson is governed by the polarization state of the $\gamma^*/Z$ boson. In amplitudes at higher order than tree level, initial-state QCD interactions of the colliding partons impart transverse momentum, relative to the collision axis, to the $\gamma^*/Z$ boson. This affects the polarization states.

In the laboratory frame, the $p\bar{p}$ collision axis is the $z$ axis, with the positive $z$ axis oriented along the direction of the proton. The transverse component of any vector, such as the momentum vector, is defined to be relative to the $z$ axis. The transverse component of vectors in other reference frames is defined to be relative to the $z$ axis in those frames.

The polar and azimuthal angles of the $\ell^-$ direction in the rest frame of the boson are denoted as $\vartheta$ and $\varphi$, respectively. For this analysis, the ideal positive $z$ axis coincides with the direction of the incoming quark so that the defintion of $\vartheta$ parallels the definition used in $e^+e^-$ collisions at LEP [7]. This frame is approximated by the Collins-Soper (CS) rest frame [10] for $p\bar{p}$ collisions. The rest frame is reached from the laboratory frame via two Lorentz boosts, first along the laboratory $z$ axis into a frame where the $z$ component of the lepton-pair momentum vector is zero, followed by a boost along the transverse component of the lepton-pair momentum vector. Within the CS frame, the $z$ axis for the polar angle is the angular bisector between the proton direction and the reverse of the antiproton direction. The positive $x$ axis for the azimuthal angle is along the direction of the transverse boost. A view of the CS frame is shown in Fig. 1. By construction, the CS-frame angles $\vartheta$ and $\varphi$ are invariant with respect to boosts along the $p\bar{p}$ collision axis. When the transverse momentum of the lepton pair is zero, the CS and laboratory coordinate-system axes are the same, and the $z$ axis and quark directions coincide if the incoming quark of the Drell-Yan parton amplitude is from the proton.

The general structure of the Drell-Yan lepton angular distribution in the boson rest frame consists of nine

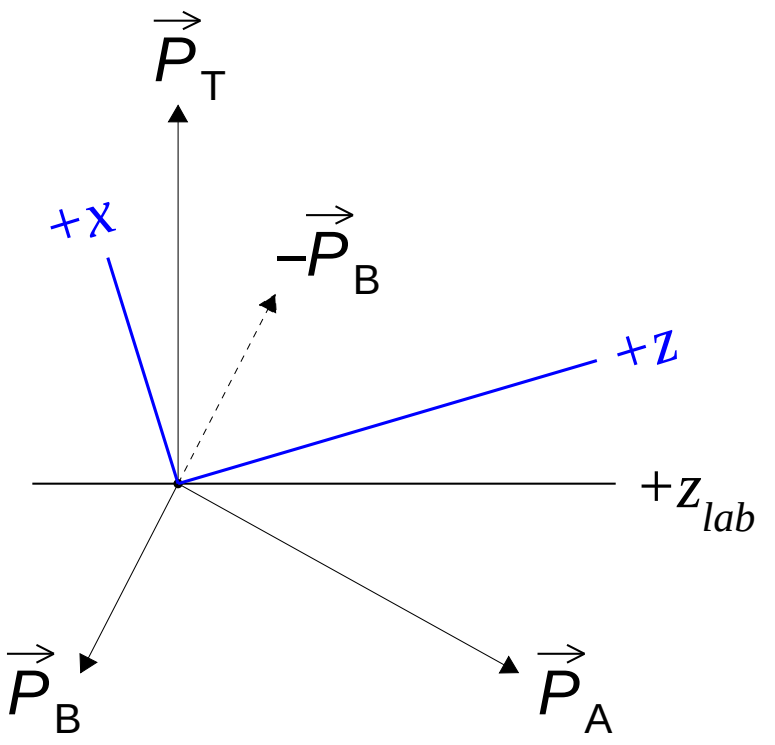

FIG. 1. Collins-Soper coordinate axes $(x, z)$ in the lepton-pair rest frame, along with the laboratory $z$ axis ($z_{lab}$). The three axes are in the plane formed by the proton ($\vec{P}_{\rm A}$) and antiproton ($\vec{P}_{\rm B}$) momentum vectors within the rest frame. Relative to the laboratory $z$ axis, the transverse component of $-(\vec{P}_{\rm A}+\vec{P}_{\rm B})$ is the same as the transverse-momentum vector of the lepton pair in the laboratory ($\vec{P}_{\rm T}$).

helicity cross-section ratios [11]:

$$\begin{aligned}\frac{dN}{d\Omega} \propto\ & (1+\cos^2\vartheta) + \\ & A_0\ \frac{1}{2}\,(1-3\cos^2\vartheta) + \\ & A_1\ \sin 2\vartheta\cos\varphi + \\ & A_2\ \frac{1}{2}\ \sin^2\vartheta\cos 2\varphi + \\ & A_3\ \sin\vartheta\cos\varphi + \\ & A_4\ \cos\vartheta + \\ & A_5\ \sin^2\vartheta\sin 2\varphi + \\ & A_6\ \sin 2\vartheta\sin\varphi + \\ & A_7\ \sin\vartheta\sin\varphi\,. \end{aligned} \quad (1)$$

The $A_{0-7}$ coefficients are the ratios of the helicity cross sections for boson production relative to unpolarized production, and are functions of kinematic variables of the boson. They vanish when the lepton-pair transverse momentum is zero, except for $A_4$, which is present at the tree level of QCD and generates the forward-backward $\ell^-$ asymmetry in $\cos\vartheta$. Thus, at zero transverse momentum, the angular distribution reduces to the tree-level form $1+\cos^2\vartheta + A_4\cos\vartheta$. The $A_4$ coefficient is relatively uniform across the range of transverse momentum where the cross section is large (under $\sim 45$ GeV/$c$), but slowly drops for larger values of transverse momentum where the cross section is very small. The $A_{5-7}$ coefficients appear at second order in the QCD strong coupling, $\alpha_s$, and are small in the CS frame [11]. Hereafter, the angles ($\vartheta$, $\varphi$) and the angular coefficients $A_{0-7}$ are specific to the CS rest frame.

The $A_4\cos\vartheta$ term is parity violating, and is due to the interference of the amplitudes of the vector and axial-vector currents. Its presence induces an asymmetry to the $\varphi$-integrated $\cos\vartheta$ cross section. Two sources contribute: the interference between the $Z$-boson vector and axial-vector amplitudes, and the interference between the photon vector and $Z$-boson axial-vector amplitudes. The asymmetric component from the $\gamma$-$Z$ interference cross section is proportional to $g_A^f$. The asymmetric component from $Z$-boson self-interference is proportional to a product of $g_V^f$ from the lepton and quark vertices, and thus is related to $\sin^2\theta_W$. At the Born level, this product is

$$T_3^\ell\,(1-4|Q_\ell|\sin^2\theta_W)\;T_3^q\,(1-4|Q_q|\sin^2\theta_W),$$

where $\ell$ and $q$ denote the lepton and quark, respectively. For the Drell-Yan process, the quarks are predominantly the light quarks: $u$, $d$, or $s$. The coupling factor has an enhanced sensitivity to $\sin^2\theta_W$ at the lepton-$Z$ vertex: as $\sin^2\theta_W \approx 0.223$, a 1% variation in $\sin^2\theta_W$ changes the lepton factor (containing $Q_\ell$) by about 8%, and it changes the quark factor (containing $Q_q$) by about 1.5% (0.4%) for the $u$ ($d$ or $s$) quark. Electroweak radiative corrections do not significantly alter this Born-level interpretation. Loop and vertex electroweak radiative corrections are multiplicative form-factor corrections to the couplings that change their value by a few percent.

For the description of the Drell-Yan process, the rapidity, transverse momentum, and mass of a particle are denoted as $y$, $P_{\rm T}$, and $M$, respectively. The energy and momentum of particles are denoted as $E$ and $P$, respectively. In a given coordinate frame, the rapidity is $y = \frac{1}{2}\,\ln[\,(E+P_{\rm z})/(E-P_{\rm z})\,]$, where $P_{\rm z}$ is the component of the momentum vector along the $z$ axis of the coordinate frame.

The $\ell^-$ forward-backward asymmetry in $\cos\vartheta$ is defined as

$$A_{\rm fb}(M) = \frac{\sigma^+(M)-\sigma^-(M)}{\sigma^+(M)+\sigma^-(M)} = \frac{3}{8}A_4(M)\,, \qquad (2)$$

where $M$ is the lepton-pair invariant mass, $\sigma^+$ is the total cross section for $\cos\vartheta \geq 0$, and $\sigma^-$ is the total cross section for $\cos\vartheta < 0$. The $\sin^2\theta^{\rm lept}_{\rm eff}$ parameter is derived from the experimental measurement of $A_{\rm fb}(M)$ and predictions of $A_{\rm fb}(M)$ for various input values of $\sin^2\theta_W$. From the prediction that best describes the measured value of $A_{\rm fb}(M)$, the value of $\sin^2\theta^{\rm lept}_{\rm eff}$ is derived. Electroweak and QCD radiative corrections are included in the predictions of $A_{\rm fb}(M)$. The QCD predictions for $A_{\rm fb}(M)$ include electroweak radiative corrections derived from an approach adopted at LEP [12].

## III. ENHANCED QCD PREDICTIONS

Drell-Yan process calculations with QCD radiation do not typically include the full electroweak radiative corrections. However, the QCD, quantum electrodynamic, and weak corrections can be organized to be individually gauge invariant so that they can be applied separately and independently.

Quantum electrodynamic (QED) radiative corrections which induce photons in the final state are not included in the calculation of $A_{\rm fb}$. Instead, they are included in the physics and detector simulation of the Drell-Yan process used in the measurement of $A_{\rm fb}$. For the process $q\bar{q} \to \ell^+\ell^-$, QED final-state radiation is most important and is included in the simulation. The effects of QED radiation are removed from the measured $A_{\rm fb}$.

The Drell-Yan process and the production of quark pairs in high-energy $e^+e^-$ collisions are analog processes: $q\bar{q} \to e^+e^-$ and $e^+e^- \to q\bar{q}$. At the Born level, the process amplitudes are of the same form except for the interchange of the electrons and quarks. Electroweak radiative corrections, calculated and extensively used for precision fits of LEP-1 and SLD measurements to the standard model [7], can be applied to the Drell-Yan process.

In the remainder of this section, the technique used to incorporate independently calculated electroweak radiative corrections for $e^+e^-$ collisions into existing QCD calculations for the Drell-Yan process is presented.

### A. Electroweak radiative corrections

The effects of virtual electroweak radiative corrections are incorporated into Drell-Yan QCD calculations via form factors for fermion-pair production in $e^+e^-$ collisions, $e^+e^- \to Z \to f\bar{f}$. The $Z$-amplitude form factors are calculated by ZFITTER 6.43 [12], which is used with LEP-1 and SLD measurement inputs for precision tests of the standard model [7]. It is a semianalytical calculation for fermion-pair production and radiative corrections for high energy $e^+e^-$ collisions. Corrections to fermion-pair production via the virtual photon include weak-interaction $W$-boson loops in the photon propagator and $Z$ propagators at fermion-photon vertices; these corrections are not gauge invariant except when combined with their gauge counterparts in the $Z$ amplitude. The ZFITTER weak and QED corrections are organized to be separately gauge invariant. Consequently, weak corrections to fermion-pair production via the virtual photon are included with the $Z$-amplitude form factors. The renormalization scheme used by ZFITTER is the on-shell scheme [13], where particle masses are on-shell, and

$$\sin^2\theta_W = 1 - M_W^2/M_Z^2 \qquad (3)$$

holds to all orders of perturbation theory by definition. Since the $Z$-boson mass is accurately known (to $\pm0.0021$ GeV/$c^2$ [7]), the inference of $\sin^2\theta_W$ is equivalent to an indirect $W$-boson mass measurement.

Form factors calculated by ZFITTER are stored for later use in QCD calculations. The specific standard model assumptions and parameters used in the form-factor calculation are presented in the appendix. The calculated form factors are $\rho_{eq}$, $\kappa_e$, $\kappa_q$, and $\kappa_{eq}$, where the label $e$ denotes an electron and $q$ denotes a quark. As the calculations use the massless-fermion approximation, the form factors only depend on the charge and weak isospin of the fermions. Consequently, the stored form factors are distinguished by three labels: $e$ (electron type), $u$ (up-quark type), and $d$ (down-quark type). The form factors are complex valued, and are functions of the $\sin^2\theta_W$ parameter and the Mandelstam $\hat{s}$ variable of the $e^+e^- \to Z \to f\bar{f}$ process. The first three form factors of the amplitude are important. They can be reformulated as corrections to the Born-level $g_A^f$ and $g_V^f$ couplings:

$$g_V^f \to \sqrt{\rho_{eq}}\,(T_3^f - 2Q_f\kappa_f\,\sin^2\theta_W) \text{ and}$$
$$g_A^f \to \sqrt{\rho_{eq}}\,T_3^f\,,$$

where $f = e$ or $q$.

The combination $\kappa_f \sin^2\theta_W$, called an effective-mixing parameter, is directly accessible from measurements of the asymmetry in the $\cos\vartheta$ distribution. However, neither the $\sin^2\theta_W$ parameter nor the form factors can be inferred from experimental measurements without assuming the standard model. The effective-mixing parameters are denoted as $\sin^2\theta_{\rm eff}$ to distinguish them from the on-shell definition of $\sin^2\theta_W$ (Eq. (3)). The Drell-Yan process is most sensitive to the parameter $\sin^2\theta_{\rm eff}$ of the lepton vertex, $\kappa_e\sin^2\theta_W$, which is commonly denoted as $\sin^2\theta_{\rm eff}^{\rm lept}$. At the $Z$ pole, $\kappa_e$ is independent of the quark flavor. For comparisons with other measurements, the value of $\sin^2\theta_{\rm eff}^{\rm lept}$ at the $Z$ pole is taken to be ${\rm Re}\,\kappa_e(\hat{s}_Z)\sin^2\theta_W$ ($\hat{s}_Z = M_Z^2$).

### B. QCD calculations

The Drell-Yan QCD calculations are improved by incorporating the form factors from ZFITTER into the process amplitude. This provides an enhanced Born approximation (EBA) to the electroweak terms of the amplitude. The QED photon self-energy correction is included as part of the EBA. The photon amplitude influences the shape of $A_{\rm fb}$ away from the $Z$ pole via its interference with the axial-vector part of the $Z$ amplitude. The $\gamma$-$Z$ interference, whose cross section is proportional to $(\hat{s} - M_Z^2)$, begins to dominate the total interference cross section away from the $Z$ pole. As the $\gamma$-$Z$ interference dilutes measurements of $\sin^2\theta_{\rm eff}$, photonic corrections are also included.

The ZFITTER form factors $\rho_{eq}$, $\kappa_e$, and $\kappa_q$ are inserted into the Born $g_A^f$ and $g_V^f$ couplings for the Drell-Yan process. The $\kappa_{eq}$ form factor is incorporated as an amplitude correction. Complex-valued form factors are used in the amplitude. Operationally, only the electroweak-coupling factors in the QCD cross sections are affected. The standard LEP $Z$-boson resonant line shape and the total decay width calculated by ZFITTER are used.

A leading-order (LO) QCD or tree-level calculation of $A_{\rm fb}$ for the process $p\bar{p} \to \gamma^*/Z \to \ell^+\ell^-$ is used as the baseline EBA calculation with ZFITTER form factors. It is used to provide a reference for the sensitivity of $A_{\rm fb}$ to QCD radiation. The CT10 [14] next-to-leading-order (NLO) parton distribution functions (PDF) provide the incoming parton flux used in all QCD calculations discussed in this section except where specified otherwise.

Two NLO calculations, RESBOS [15] and the POWHEG-BOX framework [16], are modified to be EBA-based QCD calculations. For both calculations, the boson $P_{\rm T}^2$ distribution is finite as $P_{\rm T}^2$ vanishes. The RESBOS calculation combines a NLO fixed-order calculation at high boson $P_{\rm T}$ with the Collins-Soper-Sterman resummation formalism [17] at low boson $P_{\rm T}$, which is an all-orders summation of large terms from gluon emission. The RESBOS calculation uses CTEQ6.6 [18] NLO PDFs. The POWHEG-BOX is a fully unweighted partonic-event generator that implements Drell-Yan production of $\ell^+\ell^-$ pairs at NLO. The NLO production implements a Sudakov form factor [19] that controls the infrared diverence at low $P_{\rm T}$, and is constructed to be interfaced with parton showering to avoid double counting. The PYTHIA 6.41 [20] parton-showering algorithm is used to produce the final hadron-level event.

The RESBOS and POWHEG-BOX NLO calculations are similar and consistent. The RESBOS calculation is chosen as the default EBA-based QCD calculation of $A_{\rm fb}$ with various input values of $\sin^2\theta_W$. As the POWHEG-BOX NLO program has a diverse and useful set of calculation options, it is used to estimate QCD systematic uncertainties.

## IV. THE EXPERIMENTAL APPARATUS

The CDF II apparatus is a general-purpose detector [21] at the Fermilab Tevatron $p\bar{p}$ collider whose center-of-momentum (cm) energy is 1.96 TeV. The positive $z$-axis is directed along the proton direction. For particle trajectories, the polar angle $\theta_{\rm cm}$ is relative to the proton direction and the azimuthal angle $\phi_{\rm cm}$ is oriented about the beamline axis with $\pi/2$ being vertically upwards. The component of the particle momentum transverse to the beamline is $P_{\rm T} = P\sin\theta_{\rm cm}$. The pseudorapidity of a particle trajectory is $\eta = -\ln\tan(\theta_{\rm cm}/2)$. Detector coordinates are specified as $(\eta_{\rm det}, \phi_{\rm cm})$, where $\eta_{\rm det}$ is the pseudorapidity relative to the detector center $(z = 0)$.

The central charged-particle tracking detector (tracker) is a 3.1 m long, open-cell drift chamber [22] that extends radially from 0.4 to 1.4 m. Between the Tevatron beam pipe and the central tracker is a 2 m long silicon tracker [23]. Both trackers are immersed in a 1.4 T axial magnetic field. Outside the drift chamber is a central barrel calorimeter [24, 25] that covers the region $|\eta_{\rm det}| < 1.1$. The forward end-cap regions are covered by the end-plug calorimeters [26–28] that cover the regions $1.1 < |\eta_{\rm det}| < 3.5$.

The muon detectors are outer charged-particle trackers that are positioned behind iron hadron absorbers. The primary absorbers are the calorimeters. There are four separate detectors, denoted CMU, CMP, CMX, and BMU. The CMU muon detector [29], located just beyond the central barrel calorimeter, has a cylindrical geometry and covers the region $|\eta_{\rm det}| < 0.6$. The central calorimeter provides approximately 5.5 pion (4.6 nuclear) interaction lengths of shielding. The CMP muon detector shadows the CMU detector, covers the covers the same region, $|\eta_{\rm det}| < 0.6$, but has a rectangular geometry. There are an additional 2.3 pion interaction lengths of shielding between the CMP and CMU detectors. The CMX muon detectors cover the regions $0.6 < |\eta_{\rm det}| < 1.0$, and are located behind approximately 6.2 pion interaction lengths of shielding. The BMU muon detectors cover the forward regions $1.0 < |\eta_{\rm det}| < 1.5$, and are situated behind at least 6.2 pion interaction lengths of shielding.

## V. DATA SELECTION

The data set, collected over 2002–2011, is the full CDF Run II data set and consists of $p\bar{p}$ collisions corresponding to an integrated luminosity of 9.2 fb$^{-1}$. Section V A reports on the online selection of events (triggers) for the $A_{\rm fb}$ measurement. Section V B describes the offline selection of muon candidates, and Sec. V C describes the selection of muon pairs.

### A. Triggers

Muon candidates used in this analysis are selected from two online triggers: CMUP_18 and CMX_18 [30–33]. These selections require at least one muon candidate in the event to be in the region $|\eta_{\rm det}| < 1$. The CMUP_18 selection accepts muon candidates based on track segments reconstructed in the CMU and CMP detectors that are geometrically matched to a $P_{\rm T} > 18$ GeV/$c$ charged particle track. The CMX_18 selection accepts muon candidates with a $P_{\rm T} > 18$ GeV/$c$ charged particle track in the central tracker that is matched to a track in the CMX muon detector.

### B. Offline muon selection

The offline selection begins with a charged-particle track candidate in the central tracker. The track is extrapolated through the calorimeters and into the muon detectors for association with independent track segments reconstructed in the muon detectors. The selection is based on the quality of track-to-segment matching and energy deposition in the calorimeters. The energy deposition in the calorimeters must be consistent with that of a minimum-ionizing particle. The track-to-segment matching is applied only if the track extrapolates into a fiducial region of a muon detector. The selection criteria used [21] are stringent and result in a well reconstructed sample of muon candidates with high purity.

The categories of muon candidates with associated segments in a muon detector are denoted with the following labels: CMUP, CMU, CMP, CMX, and BMU. For the CMUP category, the track extrapolation has matching segments in both the CMU and CMP detectors. The CMU category comprises muons with a matching segment in the CMU detector only. The CMP category comprises muons with a matching segment in the CMP detector only. The muons in the CMX and BMU categories have matching segments in the CMX and BMU muon detectors, respectively.

As the coverage of the muon detectors has gaps, muon candidates without associated segments in a muon detector are also used. They consist of tracks that extrapolate into nonfiducial regions of the muon detector, and fiducial tracks without matching segments. This category is denoted as CMIO (minimum-ionizing category), and consists of muon candidates that satisfy the track-quality and minimum-ionization energy loss requirement in the calorimeters.

The acceptance for muon candidates is limited by the geometric acceptance of the central tracker, whose acceptance of tracks is uniform up to $|\eta| \approx 1.1$ but then falls rapidly and vanishes at $|\eta| \approx 1.5$. In the $|\eta| > 1.1$ region, the track quality requirements for muons in the BMU category are relaxed. However, the track quality requirements for CMIO muons that have no associated muon detector segments are kept stringent.

### C. Offline muon pair selection

Events are required to contain two muon candidates. The kinematic and fiducial acceptance region for muons and muon pairs used in the $A_{\rm fb}$ measurement are listed below.

1. Muon kinematics and fiducial criteria
   (a) $P_{\rm T} > 20$ GeV/$c$;
   (b) Muon 1: CMUP or CMX category;
   (c) Muon 2: any muon category.
2. Muon-pair criteria
   (a) Muon 1 and 2: oppositely charged;
   (b) $|y| < 1$;
   (c) Muon-pair mass $M > 40$ GeV/$c^2$.

One of the muons, denoted by "Muon 1", is a CMUP or CMX muon that is consistent with the online selection. As the second muon can belong to any one of the six muon categories, eleven muon-pair topologies are possible. Muon pairs consistent with the passage of cosmic rays through the detector are rejected [21]. The limited acceptance of the central tracker restricts the accepted rapidities ($y$) of the muon pairs. As there is limited acceptance for $|y| > 1$, the $A_{\rm fb}$ measurement is restricted to muon-pairs in the kinematic region of $|y| < 1$.

The number of events passing all requirements, after background subtraction, is 276 623. The fraction of events in each of the various muon pair topologies is summarized in Table I. As the two topologies with CMP muons are rare, they are combined. The backgrounds are from QCD and the electroweak (EWK) processes of $WW$, $WZ$, $ZZ$, $t\bar{t}$, $W$+jets, and $Z \to \tau^+\tau^-$. The QCD background is primarily from dijets in which a particle in a jet has penetrated the shielding. The high-$P_{\rm T}$ muon sources yield at least one real muon. The second muon is either a real second muon or a track that is misidentified as a muon.

TABLE I. Fraction of events after background subtraction for the various muon-pair topologies. The total number of events is 276 623.

| Muon 1 | Muon 2 | Fraction |
|---|---|---|
| CMUP | CMUP | 0.159 |
| CMUP | CMX | 0.252 |
| CMUP | CMU | 0.067 |
| CMUP | CMIO | 0.181 |
| CMUP | BMU | 0.057 |
| CMX | CMX | 0.095 |
| CMX | CMU | 0.052 |
| CMX | CMIO | 0.111 |
| CMX | BMU | 0.025 |
| CMUP+CMX | CMP | 0.002 |

The EWK-process backgrounds are derived from PYTHIA [34] samples that are processed with the CDF II detector simulation, and in which the integrated luminosity of each sample is normalized to the data. The $WW$, $WZ$, $ZZ$, and $t\bar{t}$ samples are NLO simulations. As the $W$+jets and $Z \to \tau^+\tau^-$ processes are LO simulations, the total cross section used for the calculation of the integrated luminosity includes a NLO-to-LO K-factor of 1.4. The EWK-background events that pass the selection criteria amount to 0.53% of the total sample.

The QCD backgrounds are estimated from the data with the number of same-charge muon pairs in the sample, and amount to 0.10% of the total sample. The muon-pair invariant mass distributions for the data and the backgrounds are shown in Fig. 2. The distribution of same-charge muon pairs from the data sample shown in Fig. 2 also provides a measure of muon-charge misidentification. Events in which $Z \to \mu^+\mu^-$ decays are incorrectly reconstructed as same-charge muon pairs form a $Z$-resonance peak within the same-charge mass distribution. From the distribution of same-charge muon pairs shown in Fig. 2, it is concluded that charge misdentification is negligible.

Backgrounds are subtracted in the measurement of $A_{\rm fb}$, and the method is presented in Sec. VII A.

## VI. SIGNAL SIMULATION

Drell-Yan pair production is simulated using the Monte Carlo event generator, PYTHIA [34], and CDF II detector-simulation programs. PYTHIA generates the hard, leading-order QCD interaction, $q+\bar{q} \to \gamma^*/Z$, simulates initial-state QCD radiation via its parton-shower algorithms, and generates the decay $\gamma^*/Z \to \ell^+\ell^-$. The CTEQ5L [35] nucleon parton distribution functions are used in the QCD calculations. The underlying-event and boson-$P_{\mathrm{T}}$ parameters are derived from the PYTHIA configuration AW (*i.e.*, PYTUNE 101, which is a tuning to previous CDF data) [34, 36, 37]. The generator-level $P_{\mathrm{T}}$ distribution is further adjusted so that the shape of the reconstruction-level, simulated $P_{\mathrm{T}}$ distribution matches the data.

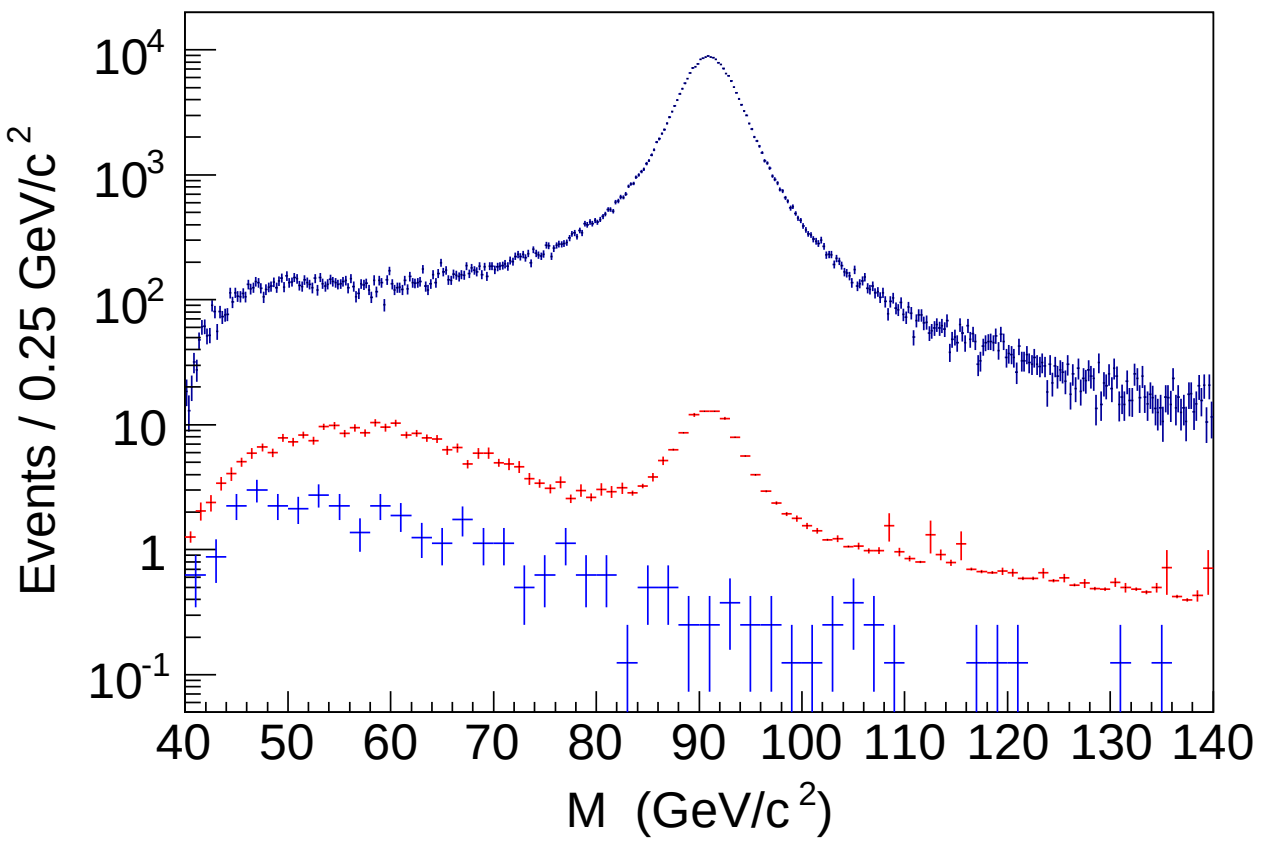

FIG. 2. Muon-pair invariant mass distributions. The upper set of crosses is the background-subtracted data, the middle set of crosses is the EWK background, and the lower set of crosses is the QCD background (same-charge muon pairs). The EWK background is derived from simulation.

Generated events are processed by the CDF II event and detector simulation. The detector simulation is based on GEANT-3 and GFLASH [38]. The event simulation includes PHOTOS 2.0 [39, 40], which adds final-state QED radiation (FSR) to decay vertices with charged particles (*e.g.*, $\gamma^*/Z \to \mu\mu$). The default implementation of PYTHIA plus PHOTOS (PYTHIA+PHOTOS) QED radiation in the simulation infrastructure has been validated in a previous measurement of $\sin^2\theta^{\rm lept}_{\rm eff}$ using Drell-Yan electron pairs [5].

The time-dependent beam and detector conditions are modeled appropriately in the simulation, including the $p$ and $\bar{p}$ beamline parameters; the luminous region profile; the instantaneous and integrated luminosities per data-taking period; and detector component calibrations, which include channel gains and malfunctions. The simulated events are reconstructed, selected, and analyzed in the same way as the experimental data.

## VII. THE $A_{\rm fb}$ MEASUREMENT

The Collins-Soper frame angle, $\cos\vartheta$ [10], is reconstructed using the following laboratory-frame quantities: the lepton energies, the lepton momenta along the beam line, the dilepton invariant mass, and the dilepton transverse momentum. The angle of the negatively charged lepton is

$$\cos\vartheta = \frac{l_+^- l_-^+ - l_-^- l_+^+}{M\sqrt{M^2 + P_{\rm T}^2}} \,,$$

where $l_\pm = (E \pm P_z)$ and the $+$ $(-)$ superscript specifies that $l_\pm$ is for the positively (negatively) charged lepton. Similarly, the Collins-Soper expression for $\varphi$ in terms of laboratory-frame quantities is

$$\tan\varphi = \frac{\sqrt{M^2 + P_{\rm T}^2}}{M} \, \frac{\vec{\Delta}\cdot\widehat{R}_{\rm T}}{\vec{\Delta}\cdot\widehat{P}_{\rm T}} \,,$$

where $\vec{\Delta}$ is the difference between the $\ell^-$ and $\ell^+$ momentum vectors; $\widehat{R}_{\rm T}$ is the transverse unit vector along $\vec{P}_p \times \vec{P}$, with $\vec{P}_p$ being the proton momentum vector and $\vec{P}$ the lepton-pair momentum vector; and $\widehat{P}_{\rm T}$ is the unit vector along the transverse component of the lepton-pair momentum vector. At $P_{\rm T} = 0$, the angular distribution is azimuthally symmetric.

The $A_{\rm fb}$ is measured in 16 mass bins, starting with $M = 50$ GeV/$c^2$. This section details the measurement method which includes corrections to the data and the simulation, and presents the fully corrected measurement. The key components of the measurement are introduced in the next two sections: Section VII A describes a newly developed event-weighting technique, and Sec. VII B describes the muon momentum and resolution calibration. Section VII C describes the data-driven corrections applied to the simulated data. Section VII D describes the resolution-unfolding technique and the corresponding covariance matrix of the unfolded $A_{\rm fb}$ measurement. Section VII E describes the final corrections to the measurement and presents the fully corrected measurement of $A_{\rm fb}$.

### A. Event-weighting method

The forward-backward asymmetry $A_{\rm fb}$ of Eq. (2) is typically determined in terms of the measured cross section $\sigma = N/(\mathcal{L}\,\epsilon A)$, where $N$ is the number of observed events after background subtraction, $\mathcal{L}$ is the integrated luminosity, $\epsilon$ the reconstruction efficiency, and $A$ the acceptance within the kinematic and fiducial restrictions. The expression is

$$A_{\rm fb} = \frac{N^+/(\epsilon A)^+ - N^-/(\epsilon A)^-}{N^+/(\epsilon A)^+ + N^-/(\epsilon A)^-} \,.$$

The terms $N^{+(-)}$ and $(\epsilon A)^{+(-)}$ respectively represent $N$ and $\epsilon A$ for candidates with $\cos\vartheta \geq 0$ ($\cos\vartheta < 0$). Each muon-pair topology listed in Table I requires a separate evaluation of $(\epsilon A)^\pm$.

The $A_{\rm fb}$ is measured using a new and simpler technique: the *event-weighting* method [41]. The method is equivalent to measurements of $A_{\rm fb}$ in $|\cos\vartheta|$ bins with these simplifying assumptions:

1. $(\epsilon A)^+ = (\epsilon A)^-$ in each $|\cos\vartheta|$ bin, and
2. Equation (1) describes the angular distributions.

The measurement of $A_{\rm fb}$ within a $|\cos\vartheta|$ bin ($A'_{\rm fb}$) only depends on $N^\pm$, but is related to $A_{\rm fb}$ through an angular dependence,

$$A'_{\rm fb} = \frac{N^+ - N^-}{N^+ + N^-} \propto A_{\rm fb} \frac{|\cos\vartheta|}{1+\cos^2\vartheta + \cdots}\,, \qquad (4)$$

where $1+\cos^2\vartheta+\cdots$ denotes symmetric terms in Eq. (1). The $|\cos\vartheta|$ term arises from the difference in the numerator $N^+-N^-$, and the $1+\cos^2\vartheta+\cdots$ term arises from the sum in the denominator $N^++N^-$. As the angular factor is the equivalent of an importance-sampling function of Monte Carlo simulations, the binned measurements are reformulated into an unbinned, event-by-event weighted expression

$$A_{\rm fb} = \frac{N_n^+ - N_n^-}{N_d^+ + N_d^-}\,. \qquad (5)$$

The $N_n^\pm$ and $N_d^\pm$ terms represent weighted event counts, and the subscripts $n$ and $d$ signify the numerator and denominator sums, respectively, which contain the same events but with different event weights. The weights take into account the angular terms of the numerator and denominator sums, and include a statistical factor for the expected measurement uncertainty at each value of $|\cos\vartheta|$, the inverse of the square of the angular factor in $A'_{\rm fb}$. Consequently, the method is equivalent to using a maximum-likelihood technique, and for an ideal detector, the statistical precision of $A_{\rm fb}$ is expected to be about 20% better relative to the direct counting method [41]. However, detector resolution and limited acceptance degrade the ideal gain.

The event weights are functions of the reconstructed kinematic variables, $\cos\vartheta$, $\varphi$, and the muon-pair variables, $M$ and $P_{\rm T}$. Only the $A_0$ and $A_2$ terms of Eq. (1) are used in the denominator of the angular factor of Eq. (4), and the angular coefficients are parametrized with

$$A_0 = A_2 = \frac{kP_{\rm T}^2}{kP_{\rm T}^2 + M^2}\,,$$

where $k$ is a tuning factor for the $P_{\rm T}$ dependence of the $A_0$ and $A_2$ coefficients. For this analysis, $k = 1.65$, which is derived from a previous measuremement of angular coefficients [42]. The exact form of these angular terms in the event weights has very little impact on $A_{\rm fb}$ because the bulk of the events is at low boson $P_{\rm T}$. The difference between $k = 1$ and $k = 1.65$ is negligible.

The EWK and QCD backgrounds are subtracted from the weighted event sums on an event-by-event basis. For the QCD same-charge pair background, $\cos\vartheta$ is calculated by randomly assigning a lepton of each pair as the negatively charged lepton. Background events passing the selection requirements are assigned negative event weights when combined with the event sums.

The event-weighting method does not compensate the following sources of bias:

1. smearing of kinematic variables due to the detector resolution,
2. kinematic regions with limited acceptance, and
3. detector nonuniformity resulting in $(\epsilon A)^+ \neq (\epsilon A)^-$.

Resolution-smearing effects are unfolded with the aid of the simulation. For the unfolding to be accurate, the muon momentum scale and resolution for both the data and simulation are precisely calibrated. In addition, the $\cos\vartheta$ and muon-pair invariant-mass distributions of the simulation are matched to agree with the data.

After resolution unfolding, the event-weighted $A_{\rm fb}$ can have a small, second-order bias. The bias is estimated using the simulation and is the difference between the true value of $A_{\rm fb}$ from the physics events generated by PYTHIA and the result of the measurement on the simulated sample. One source of bias is from the limited muon-pair acceptance at forward rapidities. There is a small increase in the forward-backward physics asymmetry with increasing $|y|$ for $|y| \approx 1$ and above. As the event-weighted $A_{\rm fb}$ is an average of $A_{\rm fb}$ over the $y$ distribution of accepted events, regions with significantly limited or no acceptance bias the average. The kinematic restriction of $|y| < 1$ specified in Sec. V C reduces this bias. Another possible source of bias is from detector nonuniformity: $(\epsilon A)^+ \neq (\epsilon A)^-$. This distorts the estimate of $A'_{\rm fb}$ [Eq. (4)]. The effects of these biases, which are quantified later in Sec. VII E, are removed from the $A_{\rm fb}$ measurement.

### B. Muon momentum calibration

The typical dependence of $A_{\rm fb}$ as a function of the lepton-pair invariant mass is shown in Fig. 3. With momentum miscalibrations, an event produced at mass $M$ with asymmetry $A_{\rm fb}(M)$ is associated with a different mass $M'$. The measured $A_{\rm fb}(M')$ becomes biased because of this systematic dilution. The correct calibration of the muon momentum is critical for the measurement of $A_{\rm fb}(M)$.

The momentum calibration procedure is adapted from a technique developed for CMS [43]. The general principles are briefly described next, followed by the CDF adaptation. The tracker is split into regions of $(\eta, \phi)$. For each region, track curvature corrections are determined. They are the curvature scale correction to the magnetic-field path integral $\int \mathbf{B}\cdot\mathbf{dl}$ and the tracking alignment offset, which are denoted by $1+s$ and $o$, respectively. The corrections $s$ and $o$ are the same for positively and negatively charged particles. For an input track curvature $C$, the corrected curvature is $(1+s)\,C + o$. In the following discussion, the curvature $C$ is synonymous to the charge-signed $1/P_{\rm T}$ of a track.

The calibration sample consists of oppositely charged muon pairs enriched in $Z$-boson decays. The muons in the sample are binned according to their $(\eta, \phi)$ trajectories. The charge-signed $1/P_{\rm T}$ for the $\mu^\pm$ is denoted by $C^\pm$, and its distribution in each bin has a sharp peak if the muon pairs are produced in the center of the tracker. The peaks become narrower as the $Z$-boson mass selection window is made smaller. The calibration method requires a single distinct peak in the $C^\pm$ distributions. The locations of these peaks are calibrated against simulated Drell-Yan muon-pair events that pass the calibration sample selection criteria. The calibration *ansatz* is that the $1 + s$ and $o$ parameters map the peaks for $C^\pm$ onto the true positions predicted by the simulation. The true location of the peaks (the truth) is the generator-level charge-signed $1/P_{\rm T}$ of the $\mu^\pm$ after QED FSR, and they are denoted by $C^\pm_{\rm true}$. Thus, the calibration constraints for $s$ and $o$ are given by

$$C^+_{\rm true} = (1+s)\,C^+ + o$$
$$C^-_{\rm true} = (1+s)\,C^- + o.$$

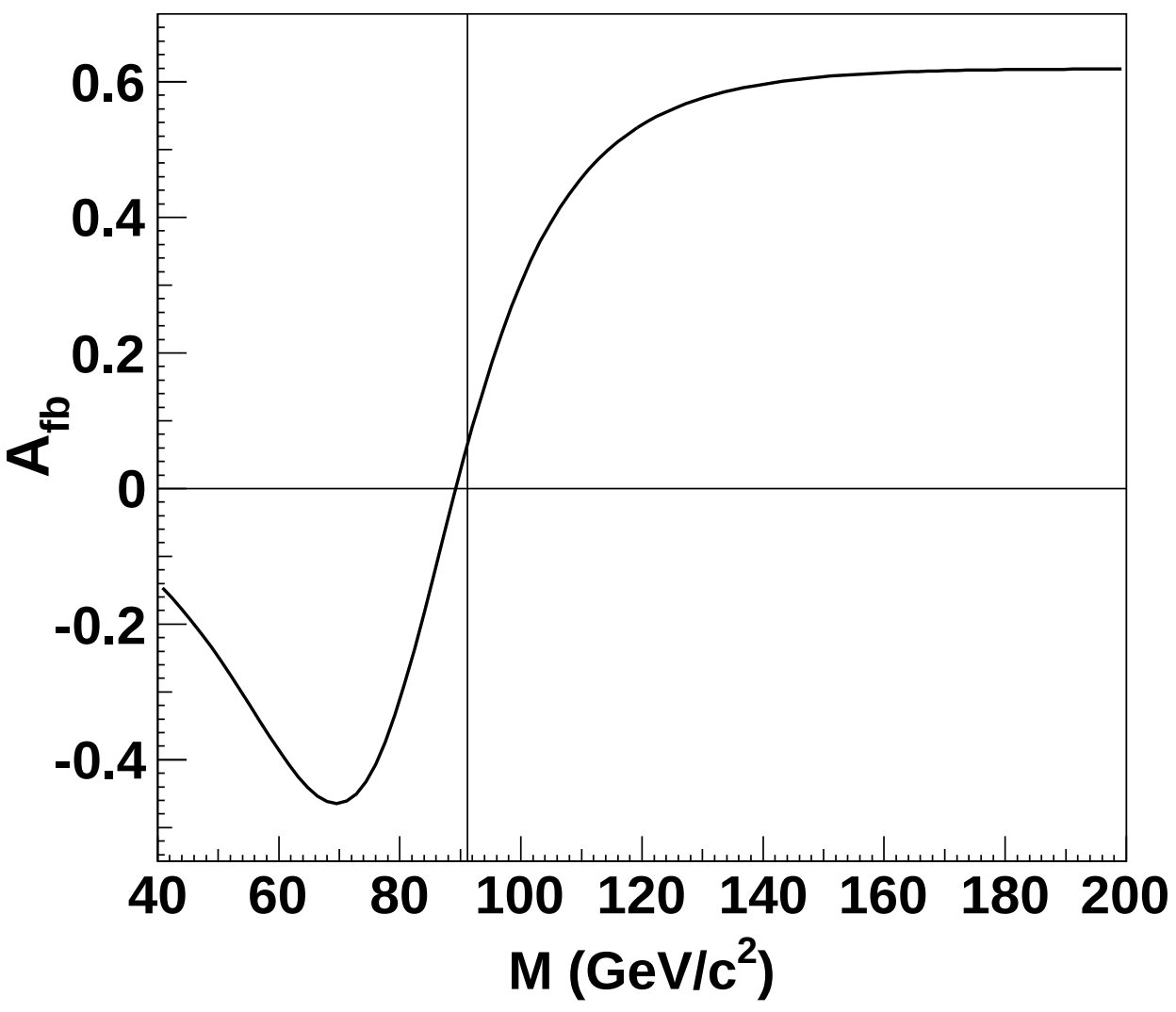

FIG. 3. Typical dependence of $A_{\rm fb}$ as a function of the lepton-pair invariant mass. The curve is an analytic calculation. The vertical line is at $M = M_Z$.

For the CDF calibration, muon pairs in the $Z$-boson region of $76 < M < 106$ GeV/$c^2$ are used. There are 262 000 events in the sample, with very little background. The muons are binned using their $(\eta, \phi)$ trajectories: eight fixed-width $\phi$ bins and eight variable-width $\eta$ bins. The $\eta$ bins span the range of $-1.6$ to 1.6, with bin boundaries of $-1.6$, $-1.0$, $-0.6$, $-0.3$, 0.0, 0.3, 0.6, 1.0, and 1.6. These bins are further divided into same-side (SS) and opposite-side (OS) muon-pair topologies: SS pairs have $\eta_1\eta_2 \geq 0$ and OS pairs have $\eta_1\eta_2 < 0$, where the subscript 1 (2) denotes Muon 1 (2). The peak of the curvature spectrum for OS-pair muons is more dependent on their point of origin along the $z$ axis than for SS-pair muons. At the Tevatron, the broad luminous region of $p\bar{p}$ collisions (30 cm longitudinal rms) has a significant impact on the higher $|\eta|$ bins. Figures 4 and 5 show the $C^+$ distribution for SS and OS pairs in one bin of the central $\eta$ region of $(-0.3, 0)$. The generator-level $C^\pm$ distributions provide an adequate description of the data for the initial steps of the iterative calibration procedure.

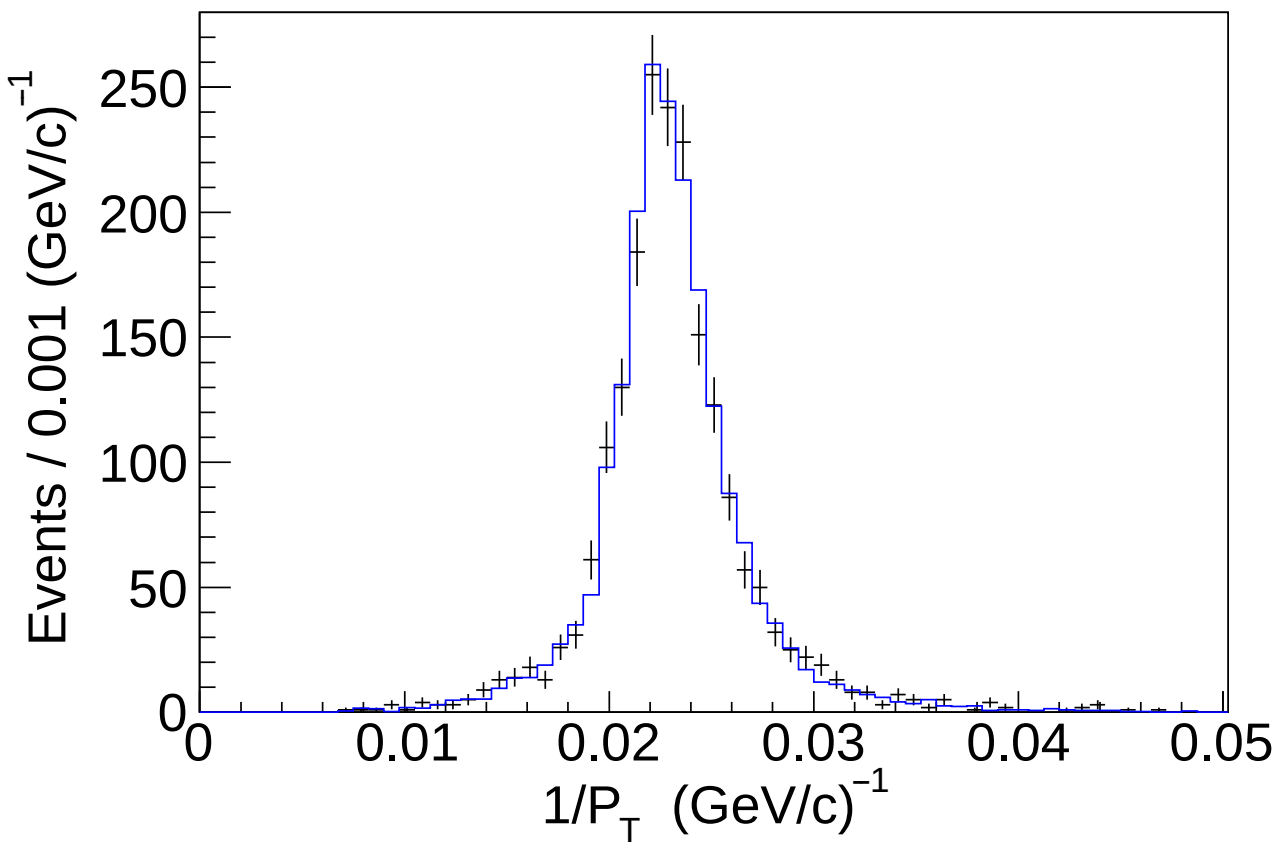

FIG. 4. Distribution of $C^+$ for SS pairs in the central $\eta$ region of $(-0.3, 0)$. The crosses are the data, and the solid histogram is the generator-level distribution normalized to the data.

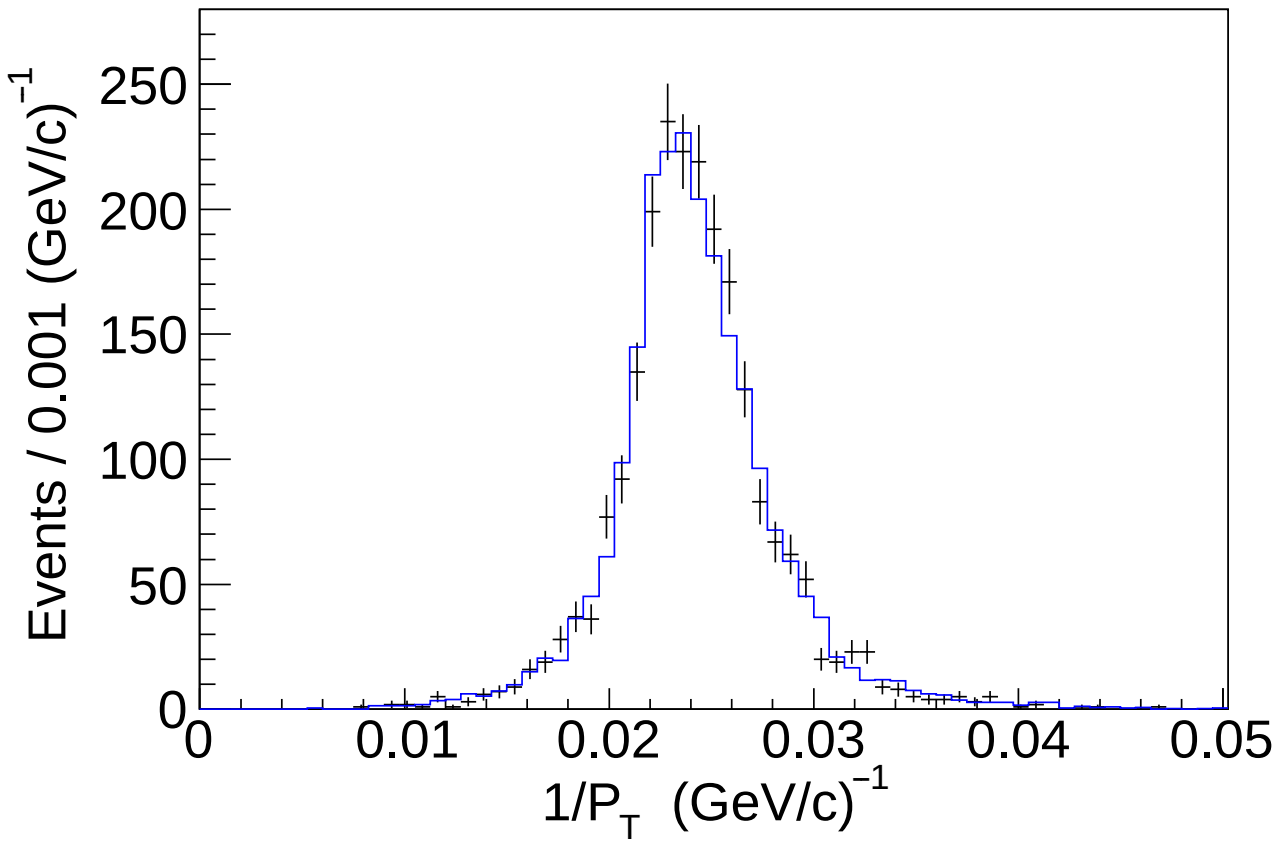

FIG. 5. Distribution of $C^+$ for OS pairs in the central $\eta$ region of $(-0.3, 0)$. The crosses are the data, and the solid histogram is the generator-level distribution normalized to the data.

The momentum scale calibration is iterative because the $s$ and $o$ calibration parameters affect the shape and location of the peaks. For the high $\eta$ bins, the calibration accuracy is no better than 1% due to the limited number of calibration events. After the third iteration with curvature peaks, the sharper $Z$-boson peak in the muon-pair invariant mass distribution is used to determine the $s$ and $o$ calibration parameters. The final three iterations use the mass peaks. For the calibration using the muon-pair invariant mass, one muon is selected as the *tag* which determines the bin. There is no bin restriction on the second muon.

The momentum scale calibration is applied to both the data and simulation. Bins that are perfectly calibrated have correction values $s = 0$ and $o = 0$. The distribution of corrections for the data is much wider than that for the simulation. In addition, corrections for the high $|\eta|$ bins are wider than those for the central region bins. For the data, the mean scale correction $s$ from the 128 calibration bins is 0.1%, and the mean alignment offset $o$ is $-0.02$ (TeV/$c$)$^{-1}$. The rms of the scale corrections is 0.4%, and the rms for the alignment offset corrections is 0.3 (TeV/$c$)$^{-1}$, or 1.4% at $P_{\rm T} = M_Z/2$. For the simulation, the mean scale correction and the mean alignment offset are 0.1% and $-0.01$ (TeV/$c$)$^{-1}$, respectively, and the corresponding rms values are 0.08% and 0.03 (TeV/$c$)$^{-1}$, respectively. The calibration of both the data and simulation sets their absolute momentum scales to the generator-level $C_{\rm true}$ scale after QED FSR.

The momentum resolution for the simulation is calibrated to the momentum resolution of the data after the scale calibrations. The resolution calibration uses the initial curvature of the simulated data, $C$. The bias of this curvature relative to its true value for each event is

$$\Delta C_{\rm true} = C_{\rm true} - C.$$

The resolution is modified by changing the amount of bias on an event-by-event basis with the parameter $f$,

$$C' = C - f\,\Delta C_{\rm true}\,,$$

where $C'$ is the new curvature. Relative to the original $C$ distribution, the rms of the $C'$ distribution is changed by the factor $1+f$. The mass distributions of muon pairs in the 86–96 GeV/$c^2$ region of the data and simulation are used to determine $f$. The value that provides the best match to the data is $f = +0.15$, and the $\chi^2$ of the simulation-to-data comparison is 68 over 79 bins.

The momentum-scale and resolution calibrations depend on the agreement between the simulated and experimental-data distributions for the $P_{\rm T}$ of the muons and invariant mass of the pair. The full results of the momentum-scale and resolution calibration are presented in the next section, which describes the data-driven corrections to the simulation.

### C. Corrections to the simulation

The simulation presented in Sec. VI does not describe the data accurately enough for the $A_{\rm fb}$ measurement. Additional corrections applied to the simulated data are described in this section. All corrections are scale factors, or event weights, that are applied to simulated events. Both the simulated and experimental data are divided into the same 39 time periods used for the offline calibration of CDF data.

The first set of corrections are event-wide corrections. The event selections described in Sec. V are applied to both the simulated and experimental data. For each muon-pair topology (Sec. V C), the number of events is adjusted period-by-period to match the data. This adjustment contains corrections to the integrated luminosity, the trigger efficiency, and global reconstruction efficiencies for each muon-pair topology. The distributions of the number of $p\bar{p}$ collision vertices in each event ($n_{\rm vtx}$) and the location of these vertices along the beamline ($z_{\rm vtx}$) changed significantly with improvements to the Tevatron collider. These distributions are inadequately simulated. The $n_{\rm vtx}$ distribution is corrected on a period-by-period basis. The $z_{\rm vtx}$ correction is split into a smaller set of seven correction blocks.

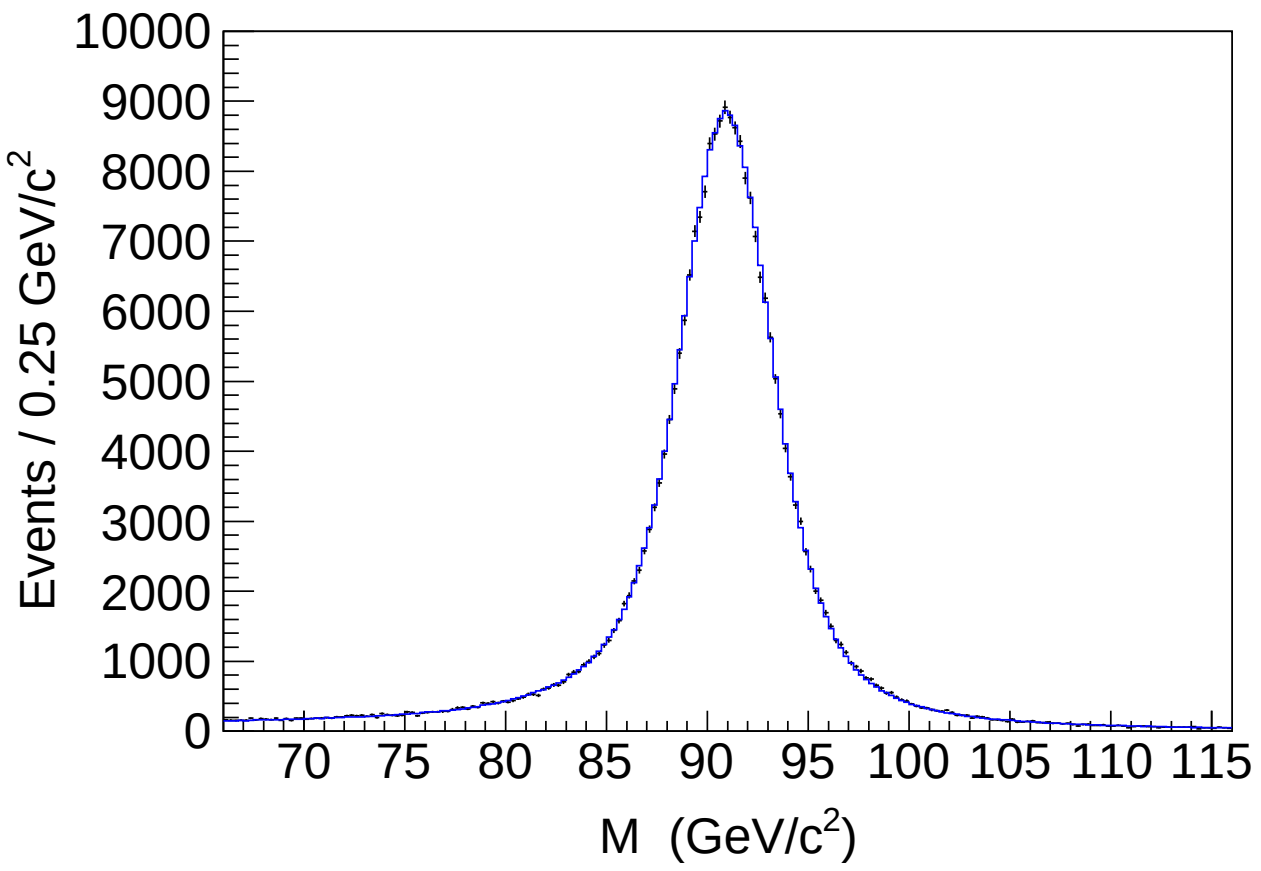

FIG. 6. Calibrated muon-pair invariant mass distributions. The crosses are the background-subtracted data and the solid histogram is from the simulation. The comparison of the simulation with the data yields a $\chi^2$ of 219 for 200 bins.

The momentum scale calibration described in the previous section is applied to both the simulated and experimental data. The momentum resolution of the simulated data is then adjusted to match the resolution of the experimental data. After these calibrations, the muon-pair invariant mass distribution of the simulated data is in good agreement with that of the experimental data. The mass distributions are shown in Figs. 6 and 7. The muon $P_{\rm T}$ distributions are shown in Figs. 8 and 9.

As the Collins-Soper $\cos\vartheta$ distribution is important for corrections to the $A_{\rm fb}$ measurement, the simulated $\cos\vartheta$ distribution is adjusted to improve agreement with the data. The adjustments, determined for eight muon-pair invariant mass bins whose boundaries are aligned with those used in the measurement, are determined from the ratios of the data-to-simulation $\cos\vartheta$ distributions. The ratios are parametrized with the function $p_0 + p_1 \cos\vartheta + p_2 \cos^2\vartheta$, where $p_0$, $p_1$, and $p_2$ are fit parameters. In the fits of the ratios with this function, the values of the asymmetry-difference parameter $p_1$ are consistent with zero. The ratios are well described by the symmetric function with $p_1 = 0$, which is used for the adjustments. The parametrized ratios are normalized to preserve the event count for the mass bin. The adjustment for the bin containing the $Z$ pole is uniform in $\cos\vartheta$. In bins away from the $Z$ pole, the adjustments redistribute events from the periphery of the $\cos\vartheta$ distribution to its center ($\cos\vartheta \approx 0$). With increasing distances of the mass bin from the $Z$ pole, the fraction of redistributed events increases, but remain under 5%. The $\cos\vartheta$ distribution after the adjustments is shown in Fig. 10. The default $\varphi$ distribution is adequate and is shown in Fig. 11.

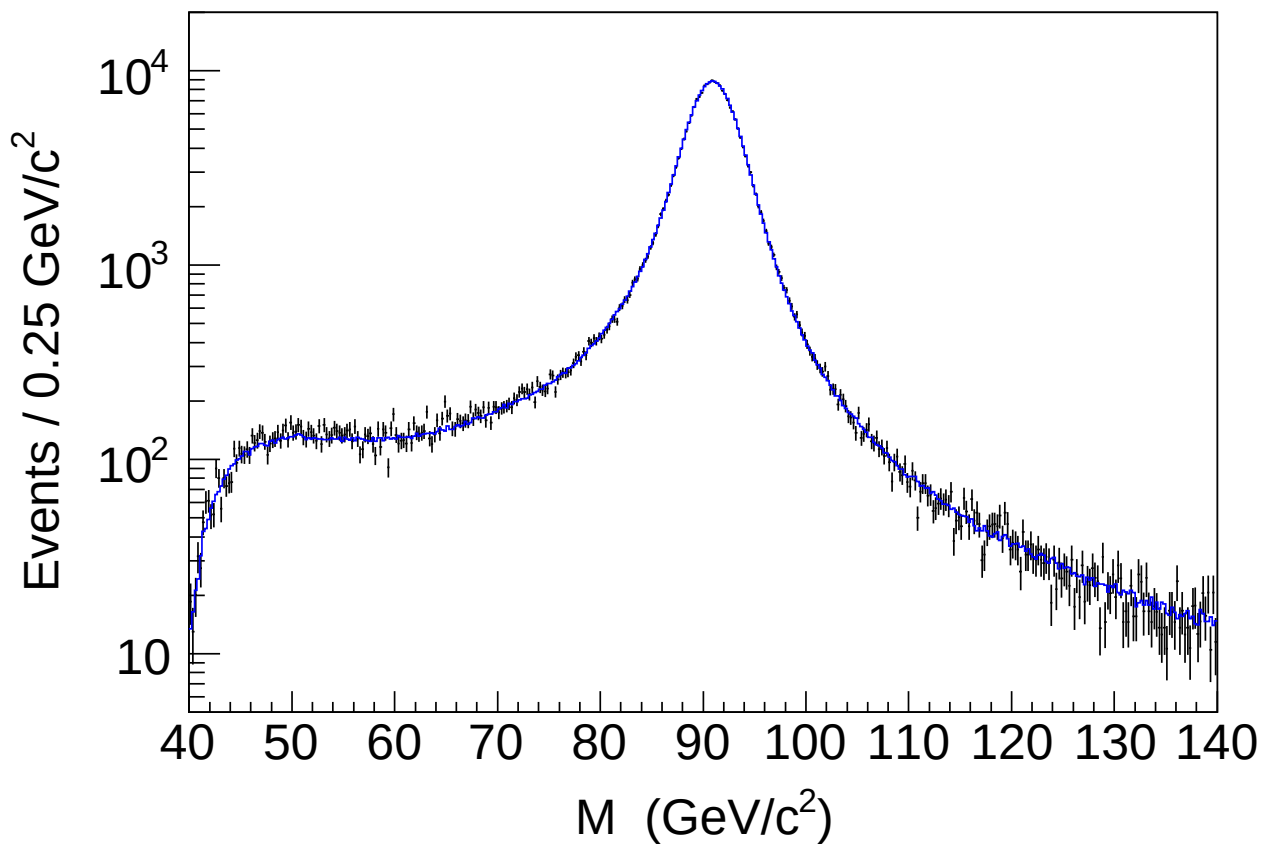

FIG. 7. Calibrated muon-pair invariant mass distributions over an extended mass range. The crosses are the background-subtracted data and the solid histogram is from the simulation. The comparison of the simulation with the data yields a $\chi^2$ of 518 for 400 bins.

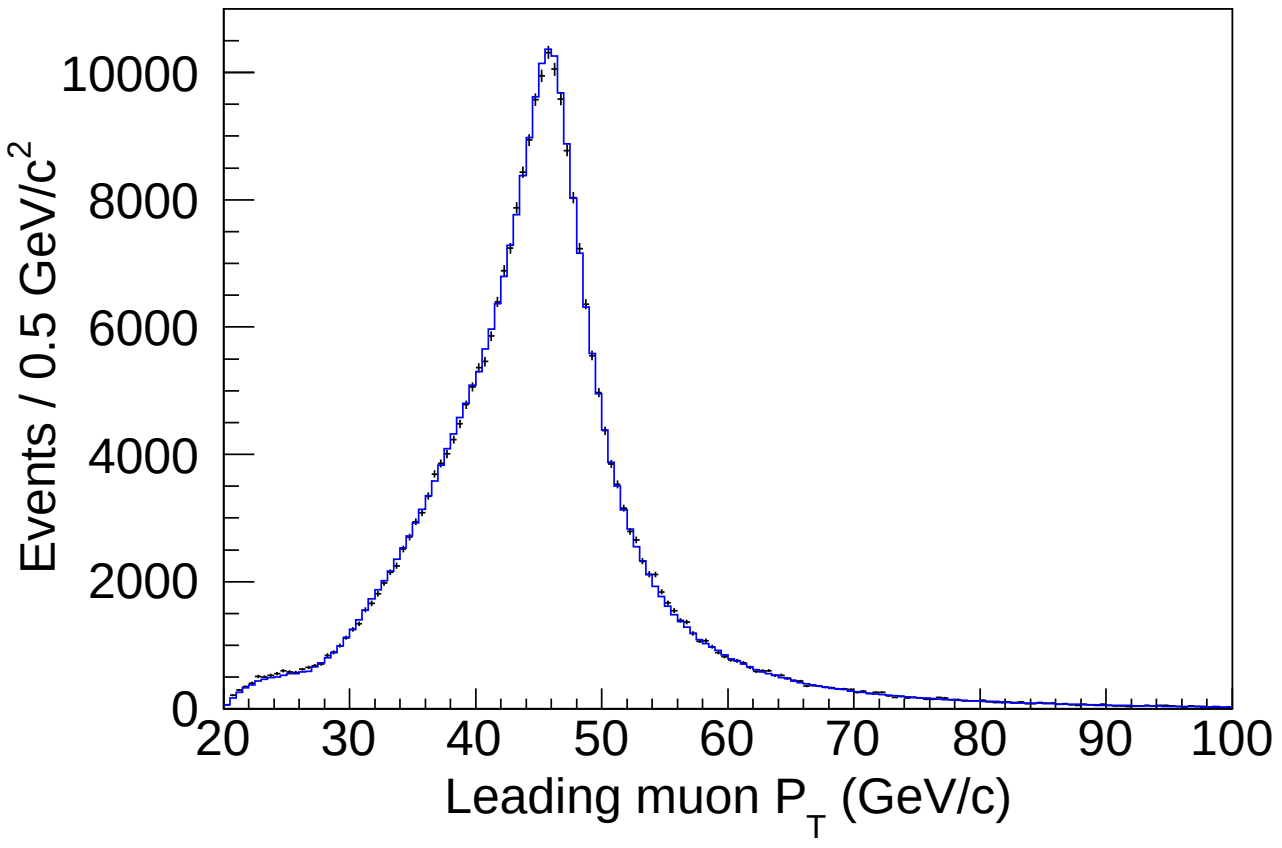

FIG. 8. Calibrated $P_{\mathrm{T}}$ distribution for the muon with the larger $P_{\mathrm{T}}$. The crosses are the background-subtracted data and the solid histogram is from the simulation.

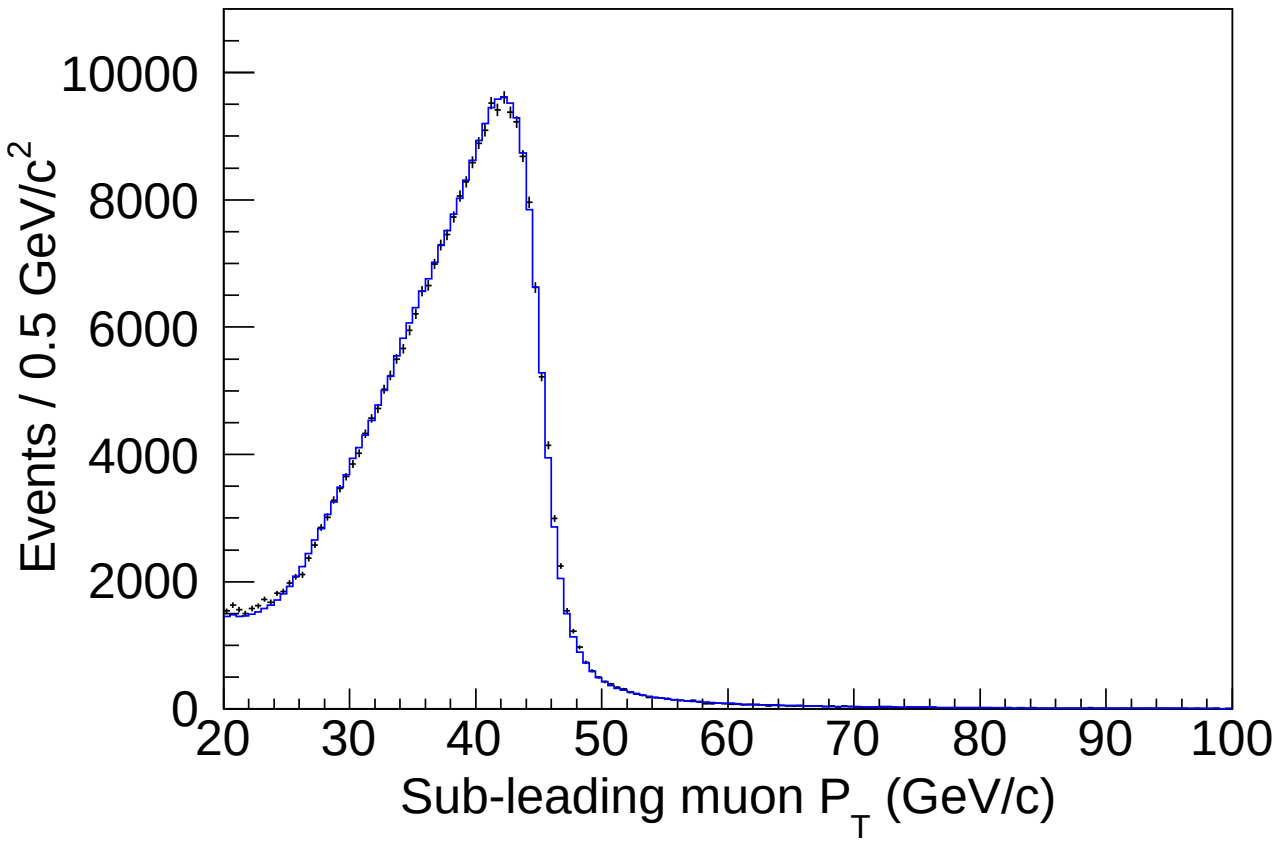

FIG. 9. Calibrated $P_{\mathrm{T}}$ distribution for the muon with the smaller $P_{\mathrm{T}}$. The crosses are the background-subtracted data and the solid histogram is from the simulation.

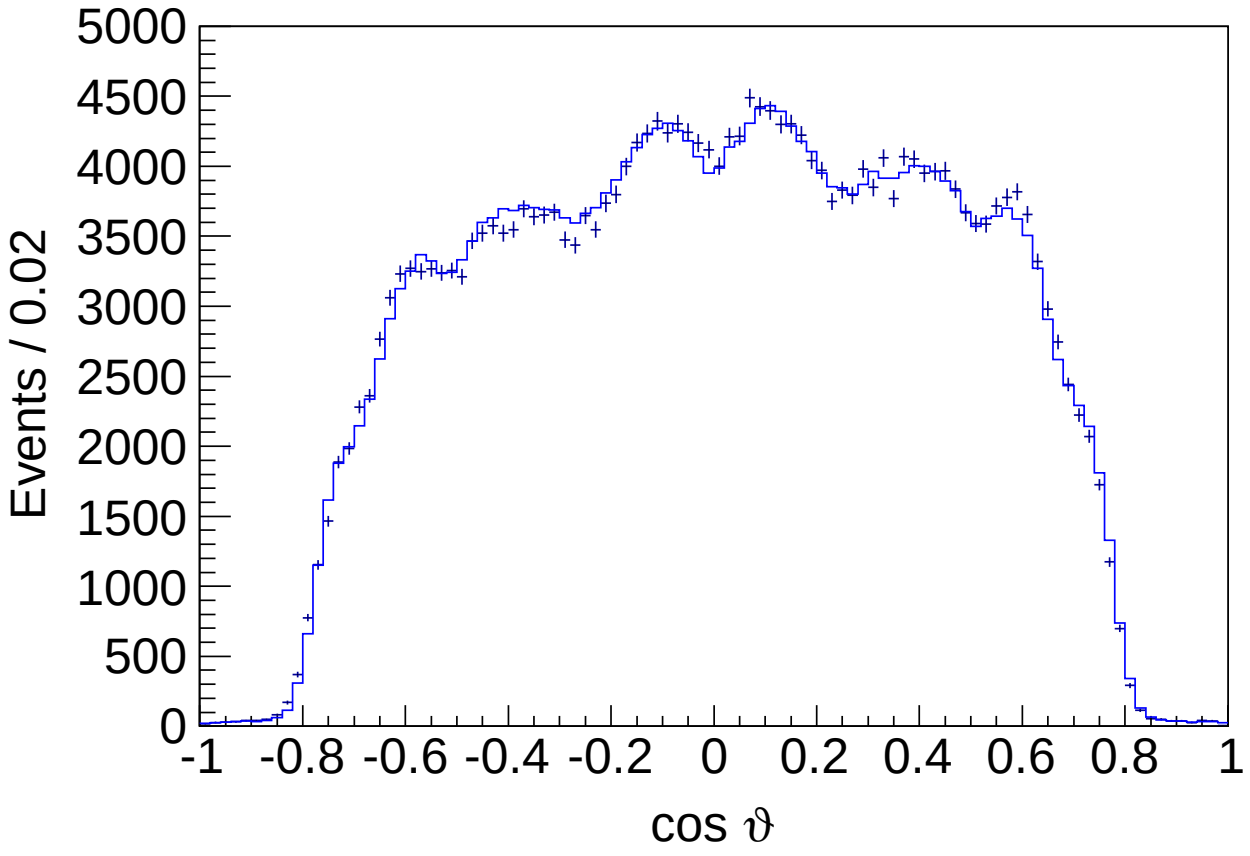

FIG. 10. Adjusted $\cos\vartheta$ distribution in the Collins-Soper frame. The crosses are the background-subtracted data and the solid histogram is from the simulation.

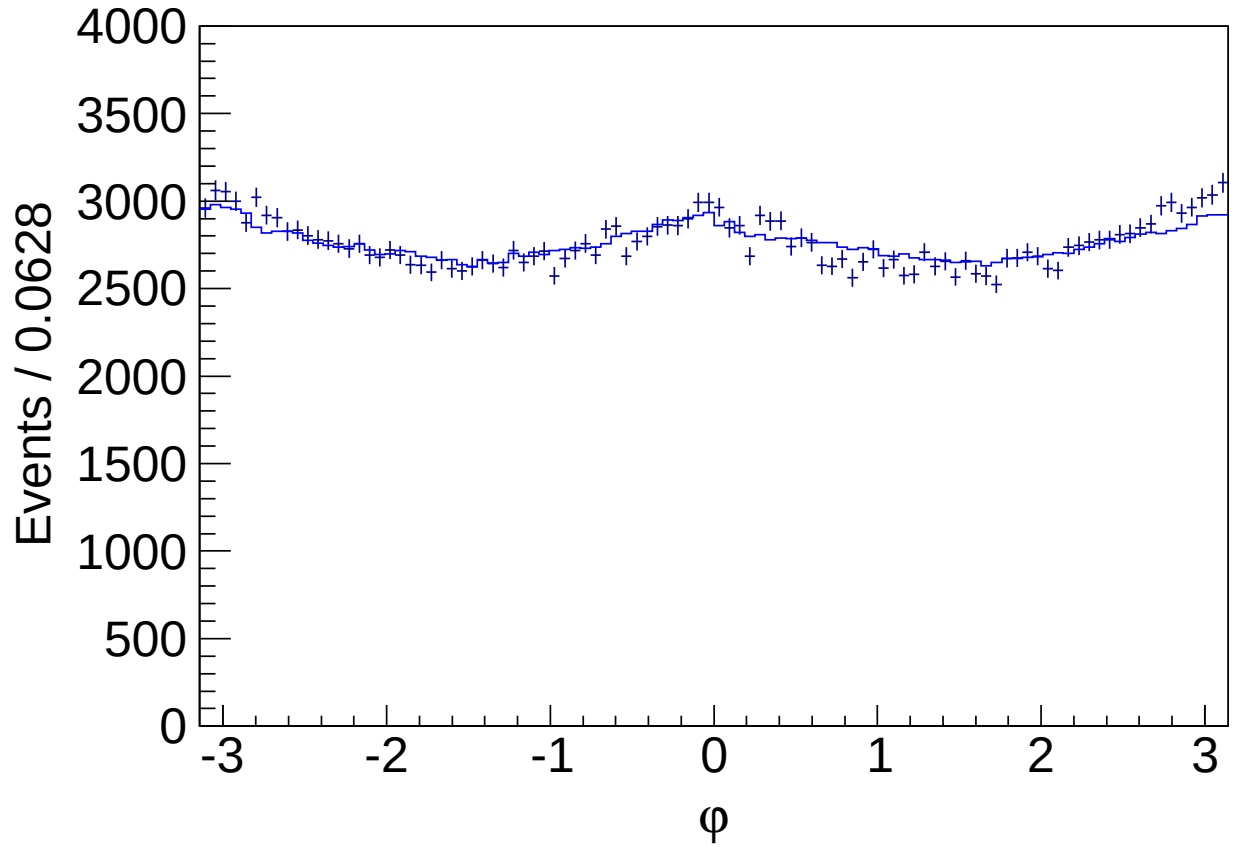

FIG. 11. Observed $\varphi$ distribution in the Collins-Soper frame. The crosses are the background-subtracted data and the solid histogram is from the simulation.

### D. Resolution unfolding

After applying the calibrations and corrections to the experimental and simulated data, the $A_{\mathrm{fb}}$ is measured in bins of the muon-pair invariant mass with the event-weighting method. This measurement is denoted as the raw $A_{\mathrm{fb}}$ measurement because the event-weighting method provides a first-order acceptance correction, but does not include resolution unfolding and final-state QED

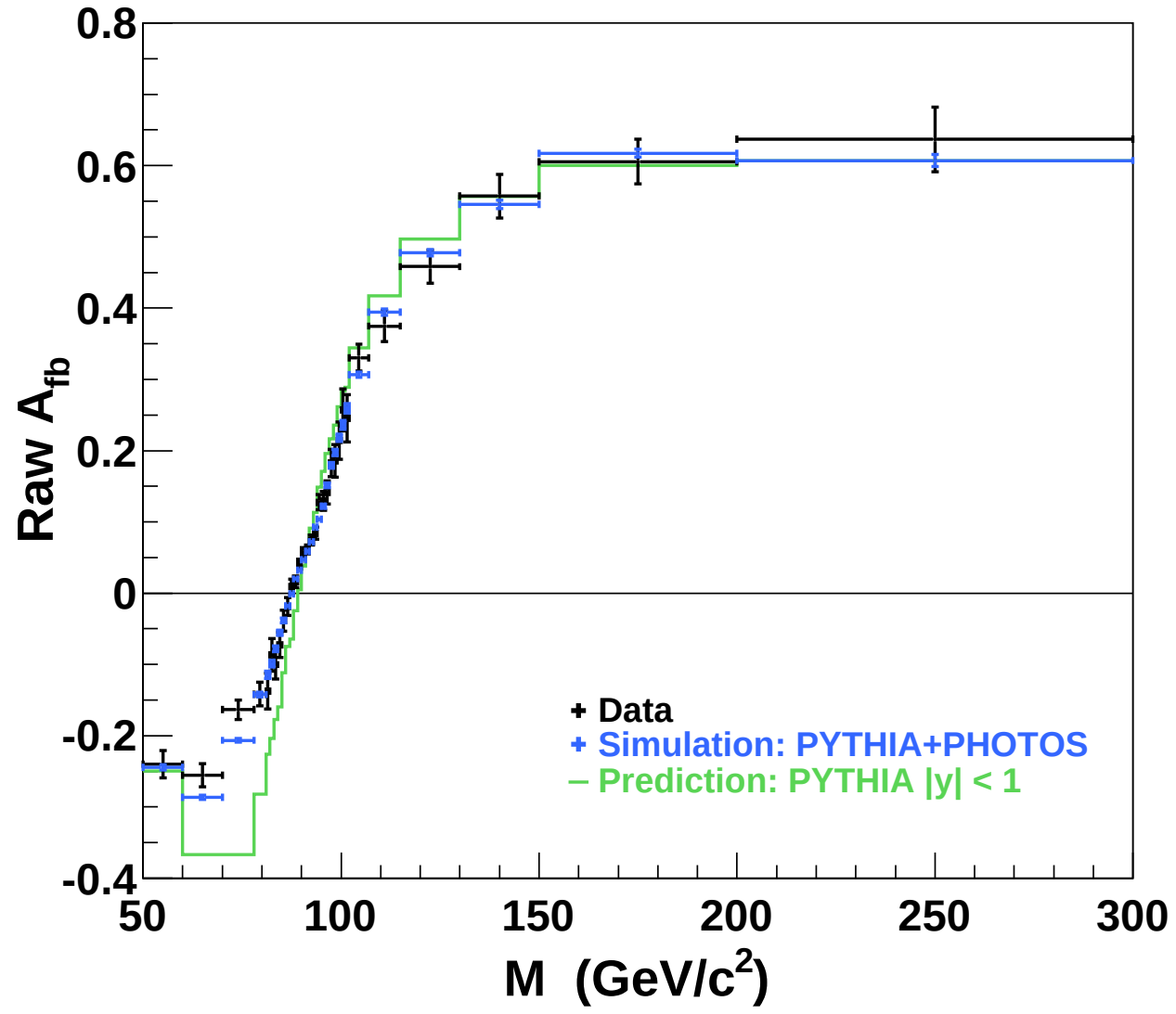

FIG. 12. Raw $A_{\rm fb}$ measurement in bins of the muon-pair invariant mass. Only statistical uncertainties are shown. The PYTHIA $|y| < 1$ asymmetry curve does not include the effect of QED FSR.

radiation. The raw $A_{\rm fb}$ measurement is shown in Fig. 12.

Resolution unfolding uses the event transfer matrices from the simulation, denoted by $\bar{n}_{gr}$. This symbol identifies the number of selected events that are generated in the muon-pair $(M, \cos\vartheta)$ bin $g$ and reconstructed in the $(M, \cos\vartheta)$ bin $r$. Sixteen mass bins are defined. Their boundaries are 50, 80, 82, 84, 86, 88, 89, 90, 91, 92, 93, 94, 96, 98, 100, 102, and 1000 GeV/$c^2$. The 50–80 and 102–1000 GeV/$c^2$ bins are referenced as the underflow and overflow bins, respectively. The forward-backward asymmetry has two angular regions $\cos\vartheta \geq 0$ $(+)$ and $\cos\vartheta < 0$ $(-)$. Operationally, $32 \times 32$ square transfer matrices for a 32-element state vector are implemented. The first 16 elements of the vector are the mass bins for the $+$ angular region, and the remaining 16 elements are for the $-$ angular region.

The simulation predicts significant bin-to-bin event migration among the mass bins when the produced and reconstructed values of $\cos\vartheta$ have the same sign. For a mass bin, there is very little migration of events from one angular region to the other. As the simulation sample size is normalized to the integrated luminosity of the data, the transfer matrices provide properly normalized estimates of event migration between bins. An estimator for the true unfolding matrix is $\bar{U}_{gr} = \bar{n}_{gr}/\bar{N}_r$, where $\bar{N}_r = \sum_g \bar{n}_{gr}$ is the expected total number of weighted events reconstructed in bin $r$. The 32-element state vector for $\bar{N}_r$ is denoted as $\vec{N}_r$, and the matrix $\bar{U}_{gr}$ by $\boldsymbol{U}$. The estimate for the resolution-unfolded state vector of produced events is $\vec{N}_g = \boldsymbol{U} \cdot \vec{N}_r$. The accuracy of the simulation of $\boldsymbol{U}$ is determined by the sample size of the data used for calibrations and corrections.

For the event-weighting method, there are two transfer matrices that correspond to the weighted event counts $N_n$ and $N_d$ of Eq. (5), and thus two separate unfolding matrices $\boldsymbol{U}$ and two separate event-weighted measurements of $\vec{N}_r$. They are used to estimate the two resolution-unfolded $\vec{N}_g$ vectors from which $A_{\rm fb}$ is derived. The measurements of $A_{\rm fb}$ for the 16 mass bins are collectively denoted by $\vec{A}_{\rm fb}$.

The covariance matrix of the $A_{\rm fb}$ measurement, denoted by $\boldsymbol{V}$, is calculated using the unfolding matrices, the expectation values of $\vec{N}_r$ and $\vec{A}_{\rm fb}$ from the simulation, and their fluctuations over an ensemble. The per-experiment fluctuation to $\vec{N}_g$ is $\boldsymbol{U} \cdot (\vec{N}_r + \delta\vec{N}_r)$, where $\delta\vec{N}_r$ represents a fluctuation from the expectation $\vec{N}_r$. The variation $\delta\vec{A}_{\rm fb}$ resulting from the $\vec{N}_g$ fluctuation is ensemble averaged to obtain the covariance matrix

$$V_{lm} = \langle\, (\delta\vec{A}_{\rm fb})_l (\delta\vec{A}_{\rm fb})_m \,\rangle \,,$$

where $(\delta\vec{A}_{\rm fb})_k$ ($k = l$ and $m$) denotes the $k$-th element of $\delta\vec{A}_{\rm fb}$. Each element $i$ of $\vec{N}_r$ receives independent, normally distributed fluctuations with a variance equal to the value expected for $\bar{N}_i$. Because $\bar{N}_i$ is a sum of event weights, fluctuations of $\bar{N}_i$ are quantified with the variance of its event weights. The two $\vec{N}_r$ vectors, the numerator vector and the denominator vector, have correlations. Elements $i$ of the numerator and denominator vectors contain the same events; the only difference being that they have different event weights. To include this correlation, the event-count variations of elements $i$ of the numerator and denominator $\delta\vec{N}_r$ vectors are based on the same fluctuation from a normal distribution with unit rms.

The covariance matrix is expanded and inverted to the error matrix using singular-value decomposition (SVD) methods. As the covariance matrix is a real-valued symmetric $16 \times 16$ matrix, its 16 eigenvalues and eigenvectors are the rank-1 matrix components in the decomposition of the covariance matrix and the error matrix

$$\boldsymbol{V} = \sum_n \lambda_n \, (\vec{v}_n \vec{v}_n) \text{ and}$$
$$\boldsymbol{V}^{-1} = \sum_n \lambda_n^{-1} \, (\vec{v}_n \vec{v}_n) \,,$$

where $\lambda_n$ and $\vec{v}_n$ are the eigenvalues and eigenvectors of $\boldsymbol{V}$, respectively, and $(\vec{v}_n \vec{v}_n)$ represents a vector projection operator, *i.e.*, $|v_n\rangle\langle v_n|$ in the style of Dirac bra-kets.

The covariance matrix has several eigenvalues with very small values. They can be interpreted as simulation noise. While they contribute very little to the structure of the covariance matrix, they completely dominate the error matrix. Consequently, comparisons between the $A_{\rm fb}$ measurement and predictions that use the error matrix are unstable. An SVD method to alleviate this instablility is used, and presented in Sec. VIII.

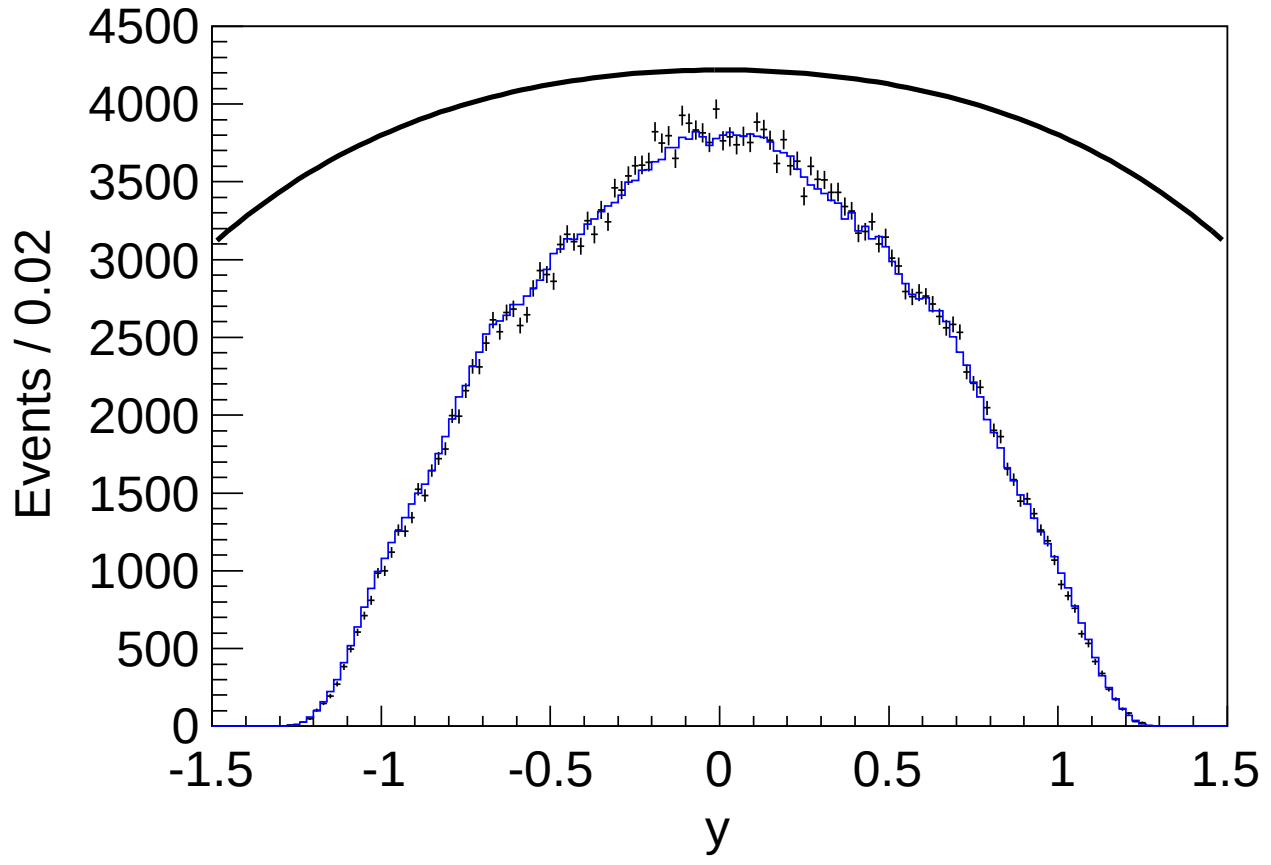

FIG. 13. Muon-pair $y$ distribution. The crosses are the background-subtracted data and the histogram is the simulation. The measurement is resticted to the region $|y| < 1$. The upper curve is the (arbitrarily normalized) shape of the underlying rapidity distribution from PYTHIA.

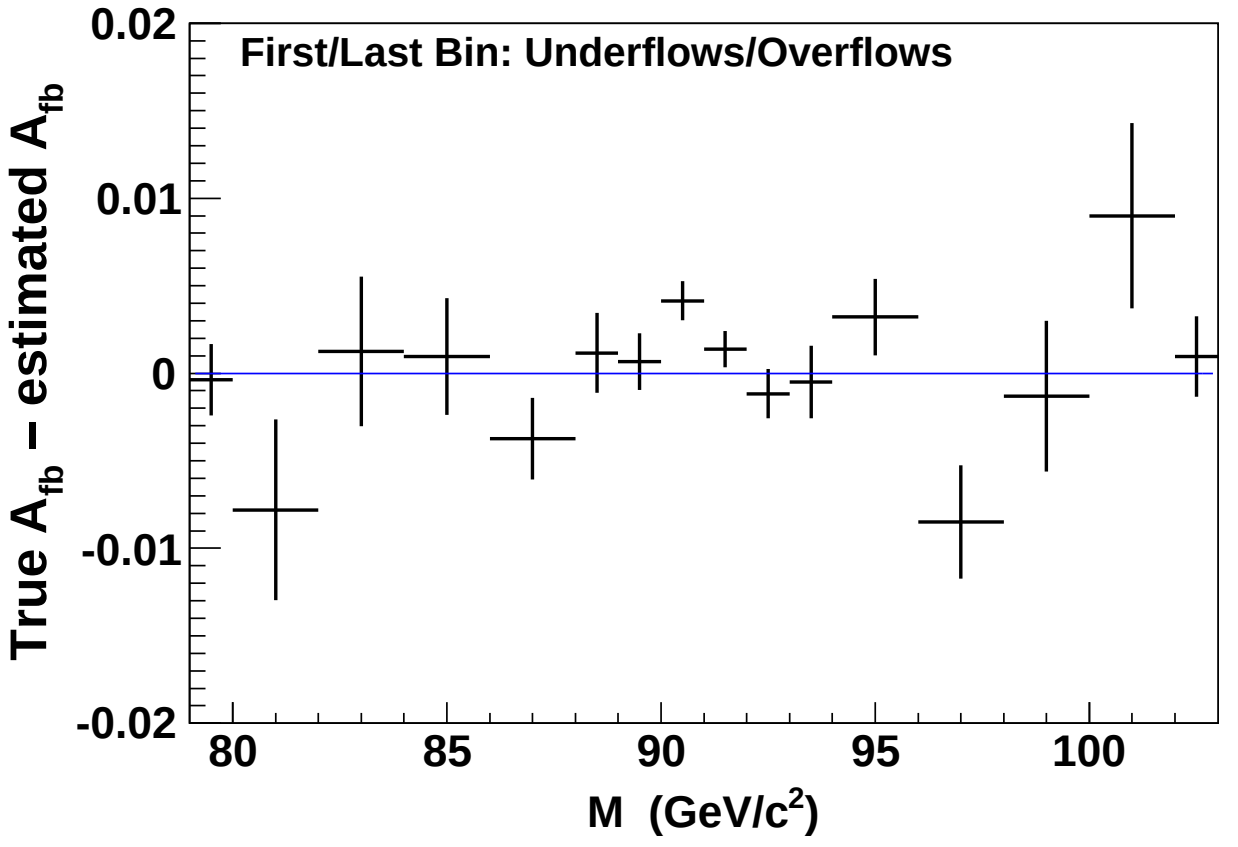

FIG. 14. Event-weighting bias for each of the muon-pair invariant mass bins. The bias is estimated with the simulation, and the uncertainties represent the full precision of the simulation.

### E. Event-weighting bias correction

After resolution unfolding, the event-weighted $A_{\rm fb}$ values can have second-order acceptance and reconstruction-efficiency biases. The most significant is the measurement bias from regions of limited boson acceptance, and to a lesser extent, from detector nonuniformities resulting in $(\epsilon A)^+ \neq (\epsilon A)^-$. The limited rapidity acceptance of muon pairs is shown in Fig. 13. As $|y|$ increases, $A_{\rm fb}$ slowly increases, and this increase is not fully taken into account in the regions of limited boson acceptance.

The bias is defined as the difference between the true value of $A_{\rm fb}$ calculated from the underlying events generated by PYTHIA and the simulation estimate. The estimate is the value of the resolution-unfolded $A_{\rm fb}$ obtained from the event-weighted simulation. Kinematic distributions of the simulated data that are important for the unfolding matrix are adjusted to agree with the data, but the adjustments exclude terms linear in the $\cos\vartheta$ kinematic variable. Linear adjustments can only be applied to the underlying physics distribution and propagated to the observed $\cos\vartheta$ distribution. The bias is a mass-bin by mass-bin additive correction to the unfolded $A_{\rm fb}$ measurement, and is shown in Fig. 14. A small net positive bias is expected due to the limited acceptance at the edges of the $|y| < 1$ measurement region for muon pairs; a bias of $(0.0009 \pm 0.0005)$ is observed. The fully corrected measurement of $A_{\rm fb}$, including the bias correction, is shown in Fig. 15 and tabulated in Table II.

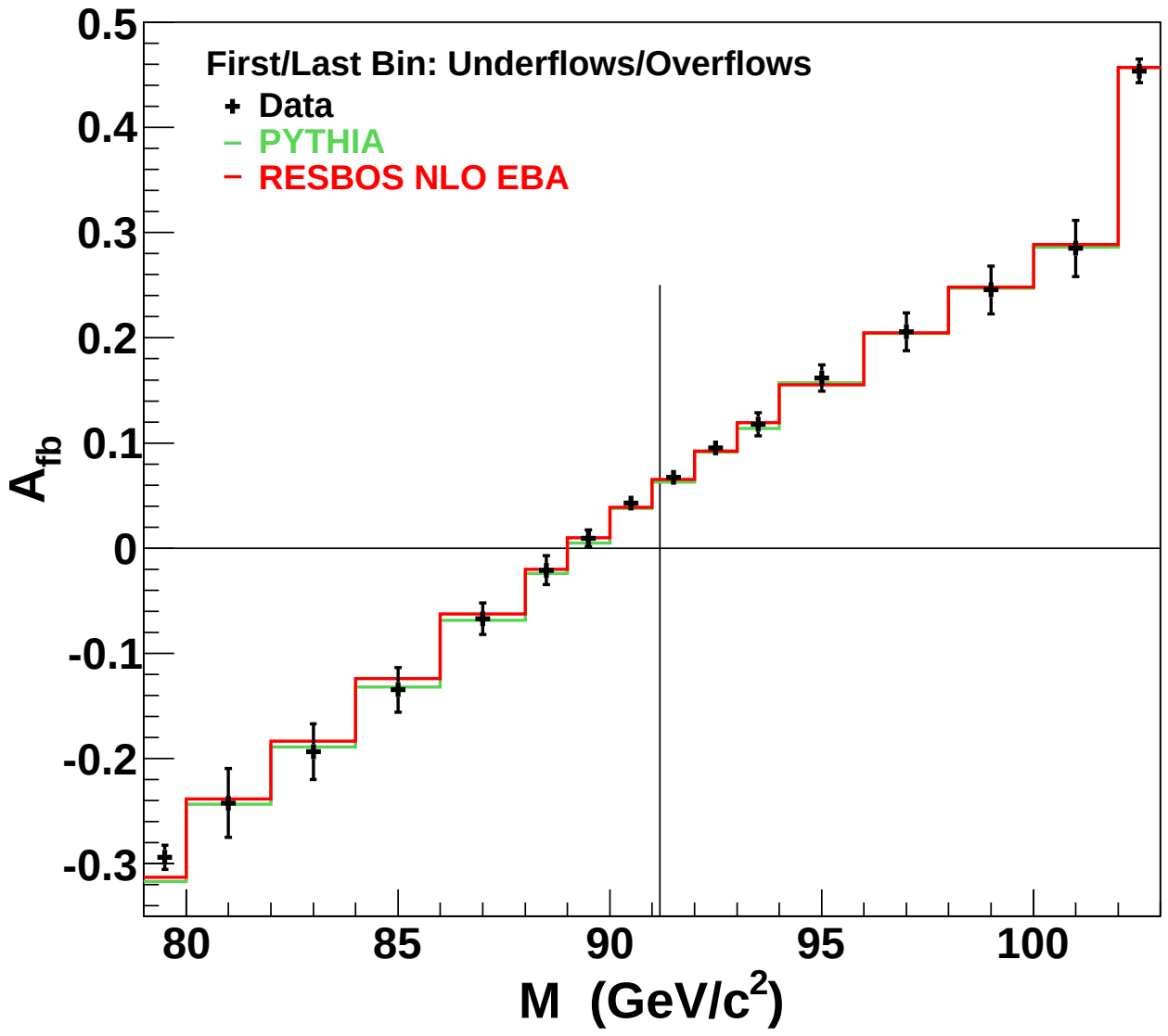

FIG. 15. Fully corrected $A_{\rm fb}$. The measurement uncertainties are uncorrelated bin-by-bin unfolding estimates. The vertical line is $M = M_Z$. The PYTHIA calculation uses $\sin^2\theta^{\rm lept}_{\rm eff} = 0.232$. The EBA-based RESBOS calculation uses $\sin^2\theta_W = 0.2233$ ($\sin^2\theta^{\rm lept}_{\rm eff} = 0.2315$).

## VIII. EXTRACTION OF $\sin^2\theta^{\rm lept}_{\rm eff}$

The EWK mixing parameters $\sin^2\theta^{\rm lept}_{\rm eff}$ and $\sin^2\theta_W$ are extracted from the $A_{\rm fb}$ measurement presented in Fig. 15 using $A_{\rm fb}$ templates calculated using different values of $\sin^2\theta_W$. Three EBA-based calculations are used: LO (tree), RESBOS NLO, and POWHEG-BOX NLO. For the EBA electroweak form-factor calculations, the weak-mixing parameter is $\sin^2\theta_W$.

The $A_{\rm fb}$ measurement is directly sensitive to the effective-mixing parameters $\sin^2\theta_{\rm eff}$, which are combinations of the form factors and $\sin^2\theta_W$ (Sec. III A). The Drell-Yan $A_{\rm fb}$ is most sensitive to the effective-leptonic $\sin^2\theta^{\rm lept}_{\rm eff}$. While the extracted values of the effective-mixing parameters are independent of the details of the EBA model, the interpretation of the best-fit value of $\sin^2\theta_W$ and its corresponding form factors are dependent on the details of the EBA model.

TABLE II. The fully corrected $A_{\rm fb}$ measurement. The measurement uncertainties are uncorrelated bin-by-bin unfolding estimates.

| Mass bin (GeV/$c^2$) | $A_{\rm fb}$ |
|---|---|
| 50–80 | $-0.294 \pm 0.011$ |
| 80–82 | $-0.242 \pm 0.033$ |
| 82–84 | $-0.194 \pm 0.027$ |
| 84–86 | $-0.135 \pm 0.021$ |
| 86–88 | $-0.067 \pm 0.015$ |
| 88–89 | $-0.021 \pm 0.014$ |
| 89–90 | $0.0093 \pm 0.0080$ |
| 90–91 | $0.0427 \pm 0.0043$ |
| 91–92 | $0.0671 \pm 0.0037$ |
| 92–93 | $0.0951 \pm 0.0062$ |
| 93–94 | $0.118 \pm 0.011$ |
| 94–96 | $0.162 \pm 0.013$ |
| 96–98 | $0.206 \pm 0.014$ |
| 98–100 | $0.246 \pm 0.023$ |
| 100–102 | $0.285 \pm 0.027$ |
| $> 102$ | $0.454 \pm 0.011$ |

The measurement and templates are compared using the $\chi^2$ statistic evaluated with the $A_{\rm fb}$ measurement error matrix. A regularization term is added to the eigenvalue coefficients of the SVD expansion of the error matrix to attenuate the contributions of noise terms with small eigenvalues. The statistical uncertainties of the bias correction and the template calculation are used as uncorrelated regularization terms. Each uncertainty is projected onto the eigenvector basis of the covariance matrix and then applied in quadrature as a regularization term:

$$\lambda_n \to \lambda_n + \sum_i (\vec{v}_n)_i^2 \Delta_i^2$$

where $\Delta_i$ is the uncertainty for mass bin $i$, and $\lambda_n$ and $\vec{v}_n$ are the eigenvalue and eigenvector, respectively, of the covariance matrix basis vector $n$. In the basis of the diagonal measurement-error matrix for $A_{\rm fb}$, these uncertainties are combined in quadrature with the measurement variance $\lambda_n$.

Each template provides a scan point for the $\chi^2$ function: $(\sin^2\theta_W, \chi^2(\sin^2\theta_W))$. The scan points are fit to a parabolic $\chi^2$ functional form:

$$\chi^2(\sin^2\theta_W) = \bar{\chi}^2 + (\sin^2\theta_W - \overline{\sin}^2\theta_W)^2/\bar{\sigma}^2 \,,$$

where $\bar{\chi}^2$, $\overline{\sin}^2\theta_W$, and $\bar{\sigma}$ are parameters. The $\overline{\sin}^2\theta_W$ parameter is the best-fit value of $\sin^2\theta_W$ and $\bar{\sigma}$ is the corresponding measurement uncertainty. The $\bar{\chi}^2$ value, relative to 16 mass bins, is the $\chi^2$ goodness-of-fit.

The $\chi^2$ distribution of the scan over templates from the RESBOS NLO calculation is shown in Fig. 16. The EBA-based RESBOS calculations of $A_{\rm fb}$ gives the central value of $\sin^2\theta_W$. The results of the template scans are summarized in Table III. Included in the table for comparison are two other measurements: the CDF 2.1 fb$^{-1}$ $ee$-pair $A_4$ result [5], and standard model $Z$-pole fits from LEP-1 and SLD [7].

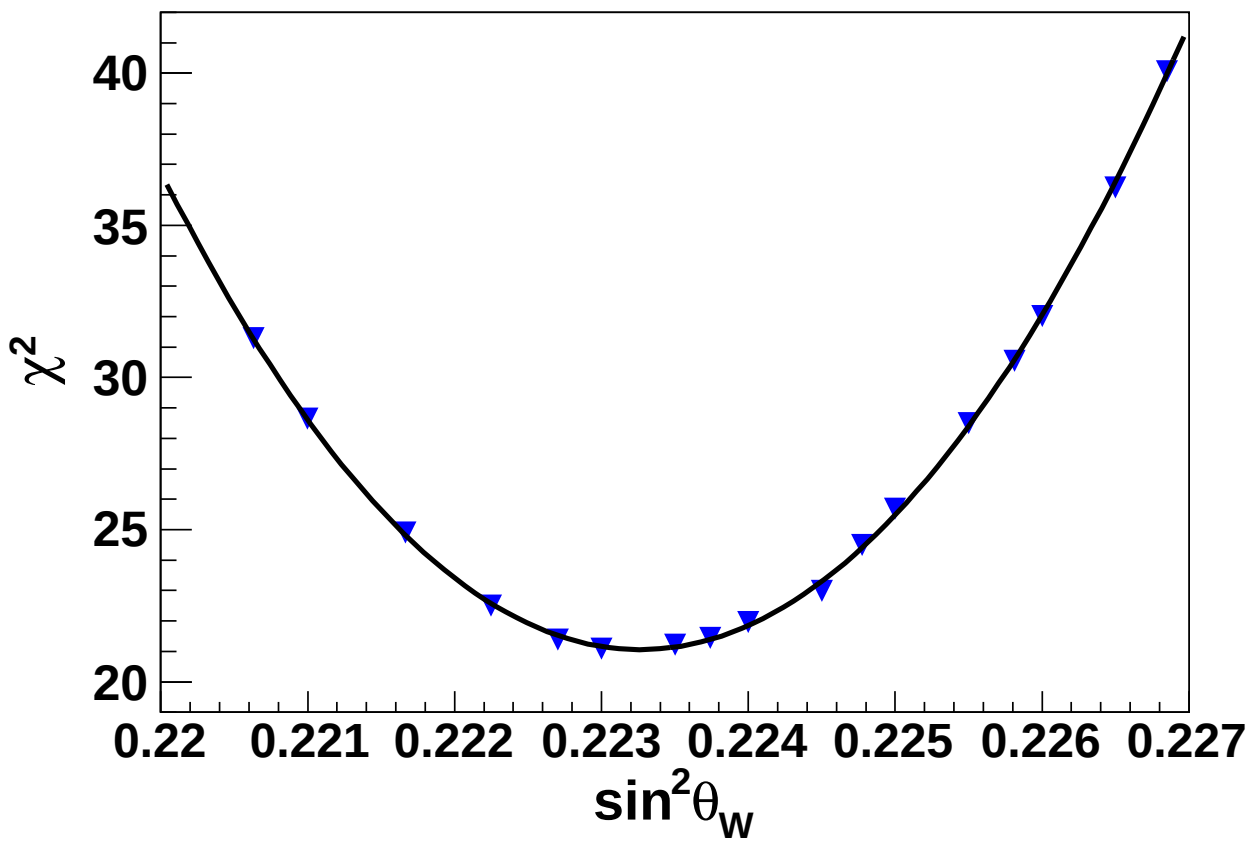

FIG. 16. Comparison of the $A_{\rm fb}$ measurement with the RESBOS NLO templates. The triangles are the scan points, and the solid curve is the fit of those points to a generic $\chi^2$ functional form.

TABLE III. Extracted values of $\sin^2\theta^{\rm lept}_{\rm eff}$ and $\sin^2\theta_W$ for the EBA-based QCD templates. The PYTHIA entry is the value from the scan over non-EBA templates calculated by PYTHIA 6.4 with CTEQ5L PDFs. The uncertainties of the template scans are the measurement uncertainties ($\bar{\sigma}$). Other measurements are listed in parentheses.

| Template (Measurement) | $\sin^2\theta^{\rm lept}_{\rm eff}$ | $\sin^2\theta_W$ | $\bar{\chi}^2$ |
|---|---|---|---|
| RESBOS NLO | $0.2315 \pm 0.0009$ | $0.2233 \pm 0.0008$ | 21.1 |
| POWHEG-BOX NLO | $0.2314 \pm 0.0009$ | $0.2231 \pm 0.0008$ | 21.4 |
| Tree LO | $0.2316 \pm 0.0008$ | $0.2234 \pm 0.0008$ | 24.2 |
| PYTHIA | $0.2311 \pm 0.0008$ | – | 20.8 |
| (CDF $A_4$) | $0.2328 \pm 0.0010$ | $0.2246 \pm 0.0009$ | – |
| (LEP-1 + SLD) | $0.23153 \pm 0.00016$ | – | – |

## IX. SYSTEMATIC UNCERTAINTIES

As the forward-backward asymmetry $A_{\rm fb}$ is a ratio of cross sections, systematic uncertainties cancel out or their effects are attenuated. The measurement of $A_{\rm fb}$ employs the event-weighting method where the simulation is used for detector resolution unfolding and the event-weighting bias correction. The level of the event-weighting bias correction is kept small by limiting the measurement of $A_{\rm fb}$ to a kinematic region where the detector acceptance is good ($|y| < 1$), and the bias correction is less than 10% of the value of $A_{\rm fb}$. The tuning of the simulation is data driven. The small residual differences from the acceptance and measurement efficiencies for the simulation relative to the data are cancelled out by the event-weighting method.

The systematic uncertainties contain contributions from both the measurement of $A_{\rm fb}$ and the template predictions of $A_{\rm fb}$ for various input values of $\sin^2\theta_W$. Both the experimental and prediction-related systematic uncertainties are small compared to the experimental statistical uncertainty. The $A_{\rm fb}$ templates from the EBA-based POWHEG-BOX calculations are used to estimate systematic uncertainties on the $\sin^2\theta_W$ parameter from various sources.

### A. Measurement

The sources investigated are muon-charge misidentification, the momentum scale, and the background estimates. Charge misidentification is found to be negligible (Sec. V C). The total measurement systematic uncertainty from the momentum scale and background is $\Delta\sin^2\theta_W = 0.00011$. The uncertainty from the backgrounds is the largest systematic uncertainty.

The reconstruction-level momentum scale of both the data and simulation are calibrated with the same technique to the underlying-physics scale. Thus, the reconstruction-level and physics-level mass bins used by the resolution unfolding and the event-weighting bias correction are aligned. However, the effect from a relative offset between the scales of the data and simulation is investigated. The global muon-momentum scale of the data is varied to determine the relative shifts allowed by the $Z$-pole mass constraint in the muon-pair invariant-mass distributions of the experimental and simulated data. The scale shift is well constrained by the precision of the data in the 66–116 GeV/$c^2$ mass range (Fig. 6). The resulting systematic uncertainty from the momentum scale is $\Delta\sin^2\theta_W = \pm 0.00005$.

Overall, the fraction of backgrounds from EWK sources is 0.53%. In the low muon-pair invariant mass region, the level is approximately 5%, and the simulated event yield in this region is slightly less than the yield of background-subtracted data. An increase in the EWK background normalization of 60% can accommodate this small difference. This normalization shift is taken as the systematic uncertainty from the background normalization, and it yields $\Delta\sin^2\theta_W = \pm 0.00010$.

### B. Predictions

The QCD mass-factorization and renormalization scales and uncertainties in the CT10 PDFs affect the $A_{\rm fb}$ templates. As the RESBOS calculation is chosen for the default $A_{\rm fb}$ templates, the associated uncertainty is also included in the overall systematic uncertainty. For the evaluation of the systematic uncertainties, the simulation equivalent of the $A_{\rm fb}$ measurement is used in template scans.

Instead of calculating the series of $A_{\rm fb}$ templates with different input values of $\sin^2\theta_W$ for each change of a QCD parameter, a simpler method is used. The $\sin^2\theta_W$ parameter is fixed to 0.2233 for all changes of QCD parameters. The predicted $A_{\rm fb}$ value for the mass bin $m$ with default QCD parameters is denoted by $\bar{A}_{\rm fb}(m,0)$, and when the QCD parameter $i$ is shifted, it is denoted by $\bar{A}_{\rm fb}(m,i)$. Each $\sin^2\theta_W$ scan point template is offset with the difference

$$A_{\rm fb}(m) \to A_{\rm fb}(m) + [\bar{A}_{\rm fb}(m,i) - \bar{A}_{\rm fb}(m,0)] \,.$$

The modified templates are then used in template scans for the best-fit value of $\sin^2\theta_W$. As there are no correlations of $A_{\rm fb}$ values among the mass bins, the simple bin-by-bin $\chi^2$ statistical measure is used for comparisons with the templates.

In all QCD calculations, the mass-factorization and renormalization scales are set to the muon-pair invariant mass. To evaluate the effects of different scales, the running scales are varied independently by a factor ranging from 0.5 to 2 in the calculations. The largest observed deviation of the best-fit value of $\sin^2\theta_W$ from the default value is considered to be the QCD-scale uncertainty. This uncertainty is $\Delta\sin^2\theta_W(\text{QCD scale}) = \pm 0.00003$.

The CT10 PDFs are derived from a global analysis of experimental data that utilizes 26 fit parameters and the associated error matrix. In addition to the best global-fit PDFs, PDFs representing the uncertainty along the eigenvectors of the error matrix are also derived. For each eigenvector $i$, a pair of PDFs are derived using 90% C.L. excursions from the best-fit parameters along its positive and negative directions. The difference between the best-fit $\sin^2\theta_W$ values obtained from the positive (negative) direction excursion PDF and the global best-fit PDF is denoted as $\delta_i^{+(-)}$. The 90% C.L. uncertainty for $\sin^2\theta_W$ is given by the expression $\frac{1}{2}\sqrt{\sum_i(|\delta_i^+| + |\delta_i^-|)^2}$, where the sum $i$ runs over the 26 eigenvectors. This value is scaled down by a factor of 1.645 for the 68.3% C.L. (one standard-deviation) uncertainty yielding $\Delta\sin^2\theta_W(\text{PDF}) = \pm 0.00036$.

The RESBOS $A_{\rm fb}$ templates are the default templates for the extraction of $\sin^2\theta^{\rm lept}_{\rm eff}$. The scan with the POWHEG-BOX or the tree templates yields slightly different values for $\sin^2\theta_W$. The difference, denoted as the EBA uncertainty, is $\Delta\sin^2\theta_W(\text{EBA}) = \pm 0.00012$. Although the RESBOS and POWHEG-BOX predictions are fixed-order NLO QCD calculations at large boson $P_{\rm T}$, they are all-orders resummation calculations in the low-to-moderate $P_{\rm T}$ region, which provides most of the total cross section. The EBA uncertainty is a combination of differences between the resummation calculations and the derived value of $\sin^2\theta_W$ with and without QCD radiation.

In summary, the total systematic uncertainties on $\sin^2\theta_W$ from the QCD mass-factorization and renormalization scales, and from the CT10 PDFs is $\pm 0.00036$. All component uncertainties are combined in quadrature.

TABLE IV. Summary of the systematic uncertainties on the extraction of the weak mixing parameters $\sin^2\theta^{\rm lept}_{\rm eff}$ and $\sin^2\theta_W$.

| Source | $\sin^2\theta^{\rm lept}_{\rm eff}$ | $\sin^2\theta_W$ |
|---|---|---|
| Momentum scale | $\pm 0.00005$ | $\pm 0.00005$ |
| Backgrounds | $\pm 0.00010$ | $\pm 0.00010$ |
| QCD scales | $\pm 0.00003$ | $\pm 0.00003$ |
| CT10 PDFs | $\pm 0.00037$ | $\pm 0.00036$ |
| EBA | $\pm 0.00012$ | $\pm 0.00012$ |

With the inclusion of the EBA uncertainty, the total prediction uncertainty is $\pm 0.00038$.

## X. RESULTS

The values for $\sin^2\theta^{\rm lept}_{\rm eff}$ and $\sin^2\theta_W$ $(M_W)$ extracted from the measurement of $A_{\rm fb}$ using $\mu^+\mu^-$ pairs from a sample corresponding to 9.2 fb$^{-1}$ are

$$\begin{aligned}\sin^2\theta^{\rm lept}_{\rm eff} &= 0.2315 \pm 0.0009 \pm 0.0004 \\ \sin^2\theta_W &= 0.2233 \pm 0.0008 \pm 0.0004 \\ M_W\ ({\rm indirect}) &= 80.365 \pm 0.043 \pm 0.019\ {\rm GeV}/c^2\ ,\end{aligned}$$

where the first contribution to the uncertainties is statistical and the second is systematic. All systematic uncertainties are combined in quadrature, and the sources and values of these uncertainties are summarized in Table IV. The inferred result on $\sin^2\theta_W$ or $M_W$ is dependent on the standard model context specified in the appendix. The $\sin^2\theta^{\rm lept}_{\rm eff}$ result is independent because of its direct relationship with $A_{\rm fb}$.

The measurement of $\sin^2\theta^{\rm lept}_{\rm eff}$ is compared with previous measurements from the Tevatron, LHC, LEP-1, and SLD in Fig. 17. The Tevatron measurements are the D0 $A_{\rm fb}$ measurement based on 5 fb$^{-1}$ of integrated luminosity [4] and the CDF measurement derived from the $A_4$ angular-distribution coefficient of $ee$-pairs from a sample corresponding to 2.1 fb$^{-1}$ of collisions [5]. The LHC measurement is the CMS analysis of Drell-Yan muon pairs from a sample corresponding to 1.1 fb$^{-1}$ of integrated luminosity [6]. The LEP-1 and SLD measurements are from measurements at the $Z$ pole. The $Z$-pole value is the combination of these six measurements

$$\begin{aligned} A^{0,\ell}_{\rm FB} &\to 0.23099 \pm 0.00053, \\ \mathcal{A}_\ell(P_\tau) &\to 0.23159 \pm 0.00041, \\ \mathcal{A}_\ell({\rm SLD}) &\to 0.23098 \pm 0.00026, \\ A^{0,{\rm b}}_{\rm FB} &\to 0.23221 \pm 0.00029, \\ A^{0,{\rm c}}_{\rm FB} &\to 0.23220 \pm 0.00081, \\ Q^{\rm had}_{\rm FB} &\to 0.2324 \pm 0.0012, \end{aligned}$$

and the light-quark value is a combination of asymmetries from the $u$, $d$, and $s$ quarks [7]. The $Q^{\rm had}_{\rm FB}$ measurement is based on the hadronic charge-asymmetry of all hadronic events.

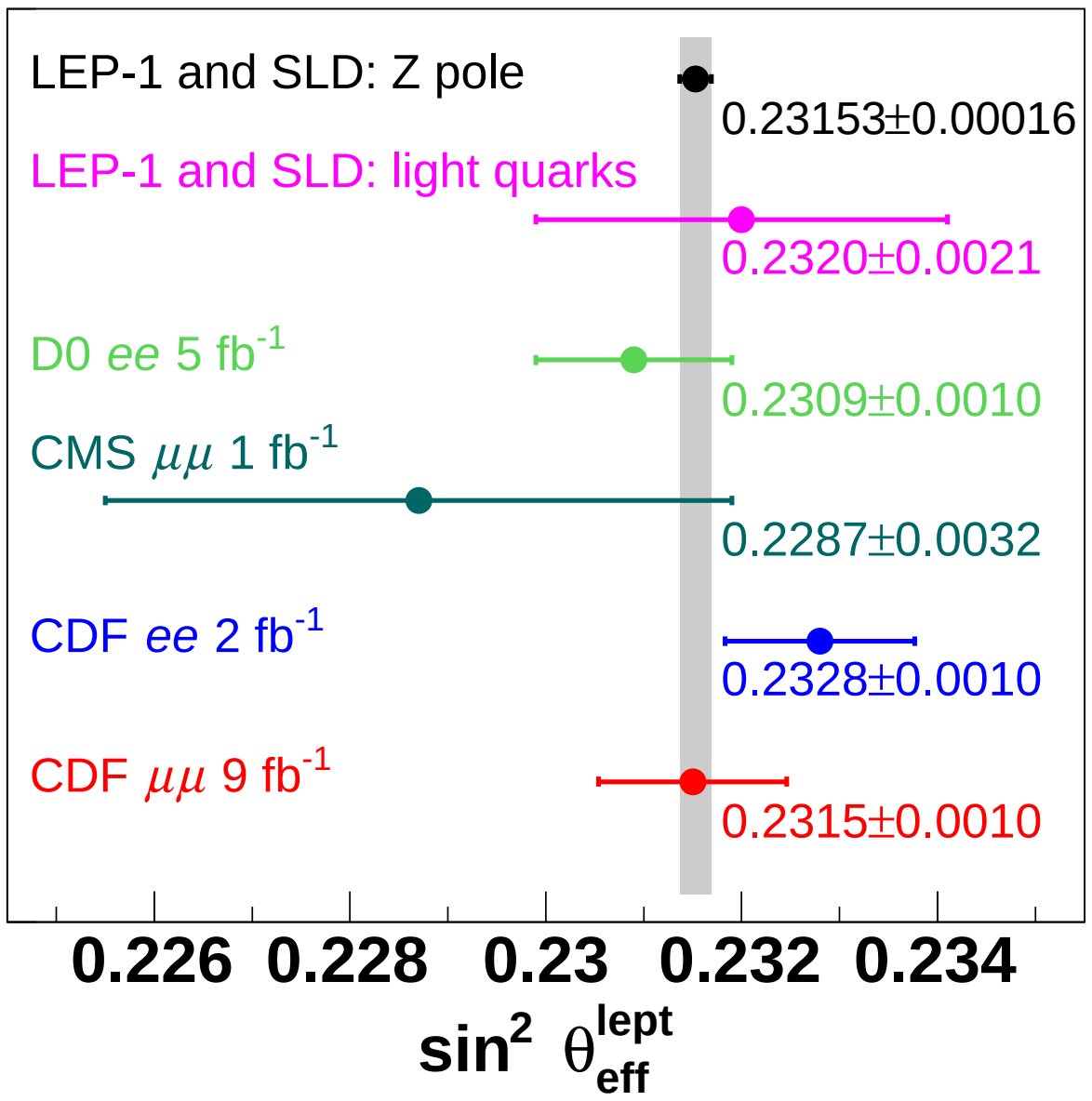

FIG. 17. Comparison of experimental measurements of $\sin^2\theta^{\rm lept}_{\rm eff}$. "$Z$ pole" represents the LEP-1 and SLD standard model analysis of $Z$-pole measurements and "light quarks" represents the LEP-1 and SLD results from the light-quark asymmetries; "D0 $ee$ 5 fb$^{-1}$" represents the D0 $A_{\rm fb}(M)$ analysis; "CMS $\mu\mu$ 1 fb$^{-1}$" represents the CMS analysis; "CDF $ee$ 2 fb$^{-1}$" represents the $A_4$ analysis; and "CDF $\mu\mu$ 9 fb$^{-1}$" represents this analysis. The horizontal bars represent total uncertainties.

The $W$-boson mass inference is compared in Fig. 18 with previous direct and indirect measurements from the Tevatron, NuTeV, LEP-1, SLD, and LEP-2. The indirect measurement from the Tevatron collider is based on the $A_4$ angular coefficient analysis [5]. The indirect measurement from LEP-1 and SLD is from electroweak standard model fits to $Z$-pole measurements in combination with the Tevatron top-quark mass measurement [44]. The NuTeV value, an indirect measurement, is based on the on-shell $\sin^2\theta_W$ parameter extracted from the measurement of the ratios of the neutral-to-charged current $\nu$ and $\bar\nu$ cross sections at Fermilab [8]. The direct measurements are from the Tevatron and LEP-2 [45]. The total uncertainties include both statistical and systematic uncertainties, which are combined in quadrature. Both CDF analyses are indirect measurements of $M_W$, and they both use the same standard model context.

## XI. SUMMARY

The angular distribution of Drell-Yan lepton pairs provides information on the electroweak-mixing parameter $\sin^2\theta_W$. The muon forward-backward asymmetry in the polar-angle distribution $\cos\vartheta$ is governed by the $A_4\cos\vartheta$ term, whose $A_4$ coefficient is directly related to the $\sin^2\theta^{\rm lept}_{\rm eff}$ mixing parameter at the lepton vertex, and indirectly to $\sin^2\theta_W$. The effective-leptonic parameter $\sin^2\theta^{\rm lept}_{\rm eff}$ is derived from the measurement of the forward-backward asymmetry $A_{\rm fb}(M)$ based on the entire CDF Run II sample of muon pairs, which corresponds to 9.2 fb$^{-1}$ of integrated luminosity from $p\bar{p}$ collisions at a center-of-momentum energy of 1.96 TeV. Calculations of $A_{\rm fb}(M)$ with different values of the electroweak-mixing parameter are compared with the measurement to determine the value of the parameter that best describes the data. The calculations include both quantum chromodynamic and electroweak radiative corrections. The best-fit values from the comparisons are summarized as follows:

$$\begin{aligned}\sin^2\theta^{\rm lept}_{\rm eff} &= 0.2315 \pm 0.0010\\ \sin^2\theta_W &= 0.2233 \pm 0.0009\\ M_W(\text{indirect}) &= 80.365 \pm 0.047 \text{ GeV}/c^2\,.\end{aligned}$$

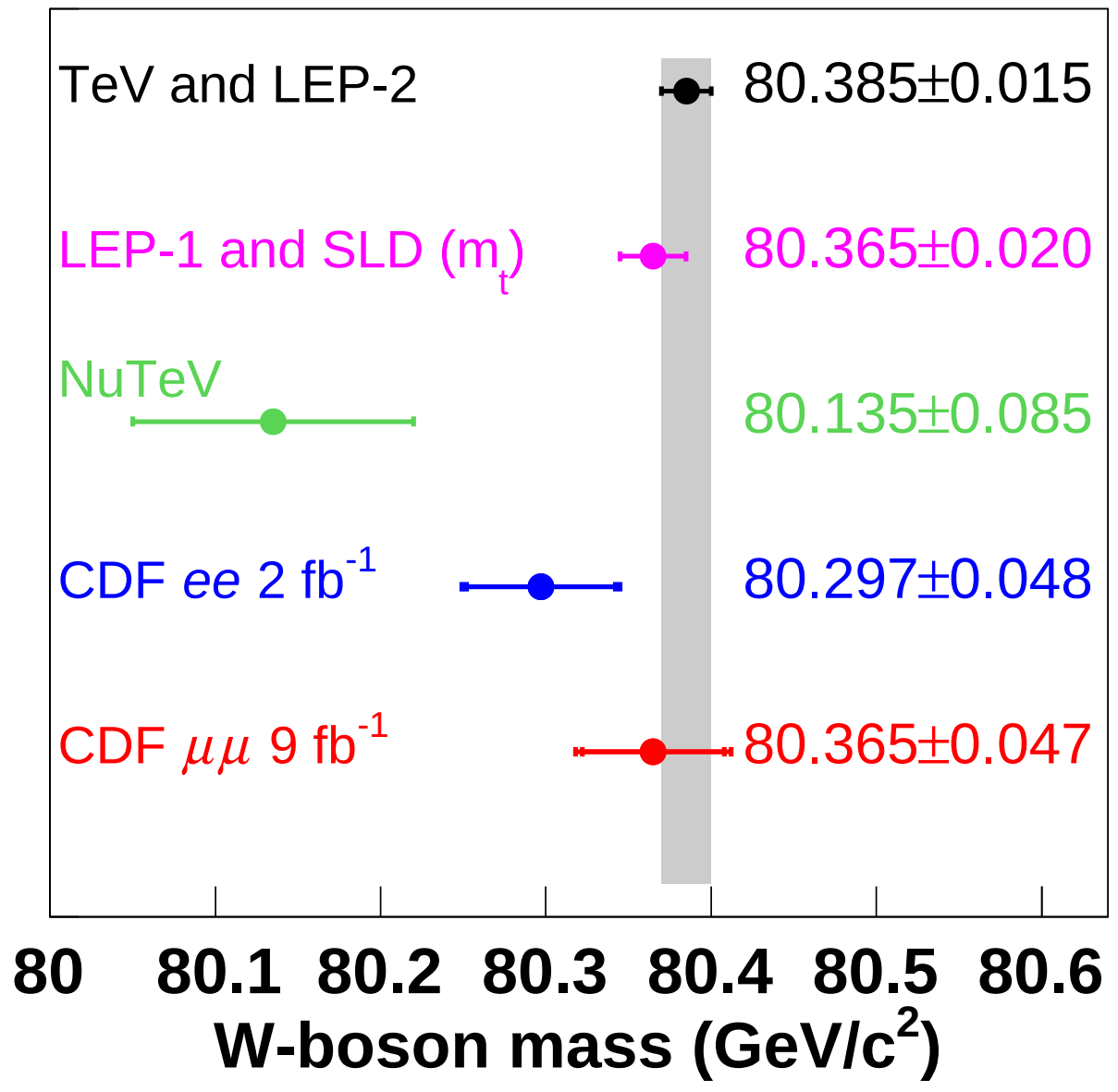

FIG. 18. Comparison of experimental determinations of the $W$-boson mass. "TeV and LEP-2" represents direct measurements of the $W$-boson mass; "LEP-1 and SLD ($m_t$)" represents the standard model analysis of $Z$-pole measurements; "NuTeV" represents the indirect measurement derived from neutrino scattering at Fermilab; "CDF $ee$ 2 fb$^{-1}$" represents the $A_4$ analysis; and "CDF $\mu\mu$ 9 fb$^{-1}$" represents this analysis. The horizontal bars represent total uncertainties.

Each uncertainty includes statistical and systematic contributions. Both results are consistent with LEP-1 and SLD measurements at the $Z$-boson pole. The value of $\sin^2\theta^{\rm lept}_{\rm eff}$ is also consistent with the previous results from the Tevatron [4, 5].

## ACKNOWLEDGMENTS

We thank the Fermilab staff and the technical staffs of the participating institutions for their vital contributions. This work was supported by the U.S. Department of Energy and National Science Foundation; the Italian Istituto Nazionale di Fisica Nucleare; the Ministry of Education, Culture, Sports, Science and Technology of Japan; the Natural Sciences and Engineering Research Council of Canada; the National Science Council of the Republic of China; the Swiss National Science Foundation; the A.P. Sloan Foundation; the Bundesministerium für Bildung und Forschung, Germany; the Korean World Class University Program, the National Research Foundation of Korea; the Science and Technology Facilities Council and the Royal Society, United Kingdom; the Russian Foundation for Basic Research; the Ministerio de Ciencia e Innovación, and Programa Consolider-Ingenio 2010, Spain; the Slovak R&D Agency; the Academy of Finland; the Australian Research Council (ARC); and the EU community Marie Curie Fellowship Contract No. 302103.

## Appendix: ZFITTER

The input parameters to the ZFITTER radiative-correction calculation are particle masses, the electromagnetic fine-structure constant $\alpha_{em}$, the Fermi constant $G_F$, the strong-interaction coupling at the $Z$ mass $\alpha_s(M_Z^2)$, and the contribution of the light quarks to the "running" $\alpha_{em}$ at the $Z$ mass $\Delta\alpha^{(5)}_{em}(M_Z^2)$ (DALH5). The scale-dependent couplings are $\alpha_s(M_Z^2) = 0.118$ and $\Delta\alpha^{(5)}_{em}(M_Z^2) = 0.0275$ [46]. The mass parameters are $M_Z = 91.1875$ GeV/$c^2$ [7], $m_t = 173.2$ GeV/$c^2$ (top quark) [44], and $m_H = 125$ GeV/$c^2$ (Higgs boson). Form factors and the $Z$-boson total decay-width $\Gamma_Z$ are calculated.

The renormalization scheme used by ZFITTER is the on-shell scheme [13], where particle masses are on-shell and

$$\sin^2\theta_W = 1 - M_W^2/M_Z^2 \qquad \text{(A.1)}$$

holds to all orders of perturbation theory by definition. If both $G_F$ and $m_H$ are specified, $\sin\theta_W$ is not independent, and it is derived from standard model constraints that use radiative corrections. To vary the $\sin\theta_W$ ($M_W$) parameter, the value of $G_F$ is changed by a small amount prior to the calculation so that the derived $M_W$ range is 80.0–80.5 GeV/$c^2$. The set of resulting $M_W$ values corresponds to a family of physics models with standard model like couplings where $\sin^2\theta_W$ and the coupling ($G_F$) are defined by the $M_W$ parameter. The Higgs-boson mass constraint $m_H = 125$ GeV/$c^2$ keeps the form factors within the vicinity of standard model fit values from LEP-1 and SLD [7].

The primary purpose of ZFITTER is to provide tables of form factors for each model. As the form factors are calculated in the massless-fermion approximation, they only depend on the fermion weak isospin and charge, and are distinguished via three indices: $e$ (electron type), $u$ (up-quark type), and $d$ (down-quark type).

For the $ee \to Z \to q\bar{q}$ process, the ZFITTER scattering-amplitude ansatz is

$$\begin{aligned} A_q = \frac{i}{4} \, & \frac{\sqrt{2} G_F M_Z^2}{\hat{s} - (M_Z^2 - i\,\hat{s}\Gamma_Z/M_Z)} \, 4T_3^e T_3^q \, \rho_{eq} \\ & [\langle \bar{e}|\gamma^\mu(1+\gamma_5)|e\rangle \langle \bar{q}|\gamma_\mu(1+\gamma_5)|q\rangle + \\ & -4|Q_e|\kappa_e \sin^2\theta_W \, \langle \bar{e}|\gamma^\mu|e\rangle \langle \bar{q}|\gamma_\mu(1+\gamma_5)|q\rangle + \\ & -4|Q_q|\kappa_q \sin^2\theta_W \, \langle \bar{e}|\gamma^\mu(1+\gamma_5)|e\rangle \langle \bar{q}|\gamma_\mu|q\rangle + \\ & 16|Q_e Q_q|\kappa_{eq} \sin^4\theta_W \langle \bar{e}|\gamma^\mu|e\rangle \langle \bar{q}|\gamma_\mu|q\rangle] \, , \end{aligned}$$

where $q = u$ or $d$, the $\rho_{eq}$, $\kappa_e$, $\kappa_q$, and $\kappa_{eq}$ are complex-valued form factors, the bilinear $\gamma$ matrix terms are covariantly contracted, and $\frac{1}{2}(1+\gamma_5)$ is the left-handed helicity projector in the ZFITTER convention. The $\kappa_e$ form factors of the $A_u$ and $A_d$ amplitudes are not equivalent; however, at $\hat{s} = M_Z^2$, they are numerically equal.

The $\rho_{eq}$, $\kappa_e$, and $\kappa_q$ form factors are incorporated into QCD calculations as corrections to the Born-level $g_A^f$ and $g_V^f$ couplings:

$$\begin{aligned} g_V^f &\to \sqrt{\rho_{eq}}\,(T_3^f - 2Q_f\kappa_f \, \sin^2\theta_W) \text{ and} \\ g_A^f &\to \sqrt{\rho_{eq}}\,T_3^f \, , \end{aligned}$$

where $f = e$ or $q$. The resulting current-current amplitude is similar to $A_q$, but the $\sin^4\theta_W$ term contains $\kappa_e\kappa_q$. The difference is removed with the addition of this amplitude correction: the $\sin^4\theta_W$ term of $A_q$ with $\kappa_{eq} \to \kappa_{eq} - \kappa_e\kappa_q$. Implementation details are provided in Ref. [5].

---